\documentclass[preprint]{article}
\usepackage{natbib}
\usepackage{url}

\usepackage{amsmath,amssymb}
\usepackage[utf8]{inputenc}
\usepackage{bbm}

\usepackage{graphicx} 
\usepackage{xcolor}
\usepackage{float}

\usepackage{soul}
\usepackage{titlesec} 
\usepackage{multirow}
\usepackage{comment}
\usepackage{tabularx}
\usepackage{longtable}
\usepackage{algorithm}
\usepackage{algpseudocode}

\newcommand{\mbf}[1]{\mathbf{#1}} 
\newcommand{\wt}[1]{\widetilde{#1}}
\newcommand{\wh}[1]{\widehat{#1}}
\newcommand{\mbm}[1]{\mathbbm{#1}} 
\newcommand{\mbb}[1]{\mathbb{#1}}

\newcommand{\mcl}[1]{\mathcal{#1}}
\newcommand{\tb}[1]{\textbf{#1}}

\def\tr{\mbox{trace}}

\newtheorem{theorem}{Theorem}
\newtheorem{lemma}{Lemma}[section]

\newtheorem{remark}{Remark}[section]
\newtheorem{definition}{Definition}[section]

\algnewcommand{\algorithmicinput}{\textbf{input:}}  
\algnewcommand{\Input}[1]{\Statex \algorithmicinput\ #1}  

\algnewcommand{\algorithmicinitial}{\textbf{initialize:}}  
\algnewcommand{\Initial}[1]{\Statex \algorithmicinitial\ #1}  

\algnewcommand{\algorithmicoutput}{\textbf{output:}}  
\algnewcommand{\Output}[1]{\Statex \algorithmicoutput\ #1}  

\newcommand{\iter}[1]{#1^{(k)}}
\newcommand{\itern}[1]{#1^{(k+1)}}
\algnewcommand{\CommentLeft}[1]{\Statex \(\triangleright\) #1}
\newcommand\independent{\protect\mathpalette{\protect\independenT}{\perp}}
\def\independenT#1#2{\mathrel{\rlap{$#1#2$}\mkern2mu{#1#2}}}

\title{Inferring Latent Graphs from Stationary Signals Using a Graphical Autoregressive Model}
  \author{Jedidiah Harwood,
    Debashis Paul, Jie Peng\footnote{Correspondence author email: jiepeng@ucdavis.edu}
    \\
    Department of Statistics, University of California,  Davis}

\date{}
\begin{document}

\maketitle

\begin{abstract}
Graphs are an intuitive way to represent relationships between variables in fields such as finance and neuroscience. However, these graphs often need to be inferred from data. In this paper, we propose a novel framework to infer a latent graph by treating the observed multidimensional data as graph-referenced stationary signals. Specifically, we introduce the \textit{graphical autoregressive model (GAR)}, where the inverse covariance matrix of the observed signals is expressed as a second-order polynomial of the normalized graph Laplacian of the latent graph. The GAR model extends the autoregressive model from time series analysis to general undirected graphs, offering a new approach to graph inference.
To estimate the latent graph, we develop a three-step procedure based on penalized maximum likelihood, supported by theoretical analysis and numerical experiments. Simulation studies and an application to S\&P 500 stock price data show that the GAR model can outperform Gaussian graphical models when it fits the observed data well. Our results suggest that the GAR model offers a promising new direction for inferring latent graphs across diverse applications. Codes and example scripts are available at \url{https://github.com/jed-harwood/SGM}.
\end{abstract}
\noindent%
\vfill

\section{Introduction}
Graphs are a powerful and intuitive tool for representing relationships between variables, such as stock prices within and across industrial sectors or brain activity across different regions. They can reveal underlying patterns in relationships that may not be evident through traditional correlation-based analyses. However, in many real-world scenarios, these graphs are not directly observable and must be inferred from the data.

Graph inference has been extensively studied in the context of structural learning for graphical models, including Gaussian graphical models (GGM) \citep{meinshausen2006high,yuan2007model,friedman2008sparse,banerjee2008model,peng2009partial,cai2011constrained}, where the goal is to infer an undirected graph whose edges represent pairwise conditional dependencies between nodes. In this paper, we propose a novel framework for inferring a latent undirected graph by modeling observed multivariate data as stationary signals on the graph.

Our approach is motivated by the observation that, in many applications, observed data can be modeled as stationary signals, with stationarity induced by a diffusion process on the underlying graph \citep{SandryhailaM2014,ThanouDKF2017,PasdeloupEtAl2018}. For example, pollution levels are influenced by geography and sources of pollution (e.g., factories), leading to spatial diffusion that can be modeled as a stationary process on a graph where nodes represent geographic locations. Similarly, international trade data, shaped by geographical proximity and political/economic agreements, can be viewed as a stationary process on a graph where countries are nodes. In neuroscience, brain activity signals, such as EEG or fMRI data, are often modeled as stationary processes on neuronal networks in the absence of external stimuli.

In this paper, we focus on the notion of stationarity defined by the co-diagonalization of the covariance matrix of the observed signals with the normalized graph Laplacian of a (latent) graph. There has been growing interest in modeling graph-referenced processes—multivariate observations where each dimension corresponds to a node in the graph—through a signal processing lens.   When the graph is known, the emphasis is on graph signal processing \citep{ShumanEtAl2013,ShumanRV2016,SandryhailaM2014,MeiM2017}. However, when the graph is unknown, the primary challenge becomes learning the latent graph from the data \citep{Kalofolias2016,DongTFV2016,EgilmezPO2017,ThanouDKF2017,segarra2017network}.

\cite{PasdeloupEtAl2018} proposed a framework where the covariance matrix of the observed signal shares the same eigenvectors as a \textit{graph-shift operator}. A graph-shift operator is a linear map that transforms a signal at a node into a linear combination of signals from neighboring nodes, with the combination coefficients determined by the adjacency matrix of the graph \citep{SandryhailaM2014}. Specifically, \cite{PasdeloupEtAl2018} represented the signal using a class of graph-shift operators called \textit{diffusion matrices}, which are symmetric matrices that satisfy specific properties, with the normalized adjacency matrix (identity matrix minus the normalized graph Laplacian) being a notable example. They characterized the solution of the unknown graph-shift operator and explored its properties. 

In a different work, \cite{segarra2017network} inferred the graph-shift operator from the eigenvectors of the covariance matrix, requiring either known or well-estimated covariance matrices.

Another line of research focuses on a related but distinct concept of regularity for graph-referenced signals by enforcing smoothness of the signal on the graph through the graph Laplacian matrix. \cite{lake2010discovering} learned the graph Laplacian by assuming that the inverse covariance of the signal equals a regularized version of the graph Laplacian. \cite{DongTFV2016} proposed a method for learning the graph Laplacian by assuming smooth graph signals and using a factor model where the covariance of the latent factors is specified as the Moore-Penrose pseudo-inverse of the graph Laplacian, and then formulating the problem as a denoising task with an $\ell_1/\ell_2$-type penalty on the graph Laplacian. \cite{Kalofolias2016} approached graph Laplacian learning by minimizing a penalized smoothness measure, using the Laplacian to measure smoothness, and placing penalties on the adjacency matrix.

In this paper, we propose a \textit{graphical autoregressive model (GAR)}, where the inverse covariance matrix of the observed signals is modeled as a second-order polynomial of the normalized graph Laplacian of a latent graph. The GAR model extends the first-order autoregressive (AR(1)) model for time series from a cyclic line graph to a general undirected graph. Our primary goal is to infer the underlying graph structure from the data, which we achieve through a three-step estimation procedure using a penalized maximum likelihood approach.

We evaluate the proposed method through theoretical analysis and extensive numerical experiments. Simulation studies and an application to S\&P 500 stock price data show that the GAR model outperforms the widely-used \textit{graphical lasso (glasso)} method \citep{friedman2008sparse} when the GAR model closely represents the underlying data structure.

The remainder of the paper is organized as follows: In Section \ref{sec:GAR_model}, we introduce the proposed GAR model. Section \ref{sec:consitency} establishes the consistency of the estimator when the dimensionality of the data diverges with the sample size at an appropriate rate. Sections \ref{sec:simulation} and \ref{sec:application} present numerical studies, including simulation experiments and a real-world application. Finally, in Section \ref{sec:discussion}, we conclude with a discussion of potential extensions of the GAR model. Additional details are provided in the appendices.

\section{Graphical Autoregressive Model}
\label{sec:GAR_model}

In this section, we first introduce the notations and define the graph Laplacian and normalized graph Laplacian matrices. We then describe the graphical autoregressive (GAR) model and propose an estimation procedure for it.

\subsection{Normalized Graph Laplacian}

Given a node set indexed by $\mathbb{V}=\{1,\cdots, p\}$, an \textit{adjacency matrix},  $\mathbb{A}=((w_{ij}))_{1 \leq i,j \leq p}$ with
$
w_{ij} \geq 0, ~w_{ij} = w_{ji}, 
$  
defines 
a \textit{weighted undirected graph}  $\mathbb{G}=\mathbb{G(V,A)}$  where an \textit{edge} $i \sim j$ for $1 \leq i \not=j \leq p$ is said to be present if and only if $w_{ij}>0$. 
Moreover, if $w_{ii}>0$, then we say that node $i$ has a \textit{self-loop} (in the following, we do not count a self-loop as an edge). 
Moreover, the \textit{degree matrix} of the graph, $\mathbb{D}=\mathbb{D(A)}$, is  a diagonal matrix with the $i$-th diagonal entry $d_i: = \sum_{j=1}^p w_{ij}$ for $i=1,\cdots, p$.

The \textit{graph Laplacian matrix} of $\mathbb{G}$ is defined as 
$\mathbb{L} := \mathbb{D}- \mathbb{A}$. 
For ease of discussion, in the following we only consider graphs without any isolated node, i.e., $d_i>0$ for $i=1,\cdots, p$. 
The \textit{normalized adjacency matrix} is defined as  $\mathbb{A}_{\mathcal{N}}:=\mathbb{D}^{-1/2} \mathbb{A} \mathbb{D}^{-1/2}$ 
and the \textit{normalized graph Laplacian matrix} is defined as $\mathbb{L}_{\mathcal{N}}  := \mathbb{D}^{-1/2} \mathbb{L} \mathbb{D}^{-1/2} = \mathbb{I}_p - \mathbb{A}_{\mathcal{N}}$, where
$\mathbb{D}^{-1/2}$ is a diagonal matrix with the $i$-th diagonal entry being $d_{i}^{-1/2}$ for $i=1,\cdots, p$, and $\mathbb{I}_p $ is the $p$ by $p$ identity matrix.

Next, we give necessary and sufficient conditions for a $p$ by $p$ matrix to be a graph Laplacian matrix and a normalized graph Laplacian matrix, respectively. 
Denote the set of $p$ by $p$ adjacency matrices for weighted undirected graphs as $\mathcal{A}_p:=\{\mathbb{A}=((w_{ij}))_{1 \leq i,j \leq p}: w_{ij} \geq 0, ~w_{ij} = w_{ji} ~ \text{for all}~ 1 \leq i,j \leq p\}$. Denote the set of   $p$ by $p$ graph Laplacian matrices as $\mathcal{L}_p:=\{\mathbb{L}=\mathbb{D(A)}-\mathbb{A}: \mathbb{A} \in \mathcal{A}_p\}$ and the set of $p$ by $p$ normalized graph Laplacian matrices as  $\mathcal{L}_{\mathcal{N},p}:=\{\mathbb{L}=\mathbb{I}_p-\mathbb{A}_{\mathcal{N}}: \mathbb{A} \in \mathcal{A}_p\}$.

\begin{lemma}
	\label{lemma: L_space}
A $p$ by $p$ matrix $\mathbb{L}$ belongs to $\mathcal{L}_p$ if and only if $\mathbb{L}$ satisfies the following:
(i) $\mathbb{L}=\mathbb{L}^T$; (ii) $\mathbb{L} \mathbf{1}_p =\mathbf{0}_p$; (iii) $\mathbb{L}_{ij} \leq 0$ for all $1 \leq i \not=j \leq p$,
where $\mathbf{1}_p$ and $\mathbf{0}_p$ denote the p-dimensional vectors of $1$s and $0$s, respectively, and $\mathbb{L}_{ij}$ denotes the $(i,j)$th entry of $\mathbb{L}$. 
In addition, if $\mathbb{L} \in \mathcal{L}_p$, then  (iv) $\mathbb{L}$ is positive semi-definite (p.s.d.) with the smallest eigenvalue $\lambda_{\min}({\mbb{L}})=0$.
\end{lemma}

\begin{lemma}
	\label{lemma: L_N_space}
A $p$ by $p$ matrix $\mathbb{L}$ belongs to $\mathcal{L}_{\mathcal{N}, p}$ if and only if $\mathbb{L}$ satisfies the following:
(i) $\mathbb{L}=\mathbb{L}^T$; (ii) there exists a vector $\mathbf{v}_0 \in \mathbb{R}^p$ with all positive entries  (denoted by $\mbf{v}_0 \succ \mbf{0}$) such that $\mathbb{L} \mathbf{v}_0 =\mathbf{0}_p$; (iii) $\mathbb{L}_{ii}\leq 1$ for all  $i=1,\cdots p$ and $\mathbb{L}_{ij} \leq 0$ for all $1 \leq i \not=j \leq p$.  In addition, if $\mathbb{L} \in \mathcal{L}_{\mathcal{N},p}$, then   (iii)'  $0 \leq \mathbb{L}_{ii}\leq 1$ for all  $i=1,\cdots p$ and $-1 \leq \mathbb{L}_{ij} \leq 0$ for all $1 \leq i \not=j \leq p$; and (iv) $\mathbb{L}$ is positive semi-definite (p.s.d.) with the smallest eigenvalue $\lambda_{\min}({\mbb{L}})=0$.
\end{lemma}

Note that the $\mbf{v}_0$ vector in Lemma \ref{lemma: L_N_space} is an eigenvector of $\mbb{L}$ associated with the zero eigenvalue. Moreover, it is easy to see that the vector of the square-root degree sequence, $(\sqrt{d_1},\cdots, \sqrt{d_p})^T$, is an eligible $\mbf{v}_0$. The proofs of these lemmas are given in Appendix \ref{sec: appendix_lemma}.  
Also note that, the adjacency matrix, the graph Laplacian matrix, the normalized adjacency matrix and the normalized graph Laplacian matrix all have the same zero-pattern on their off-diagonal entries, which corresponds to the topology (i.e., presence and absence of edges) of the graph $\mathbb{G}$.

\subsection{GAR Model}

We now introduce the \textit{graphical autoregressive (GAR) model}. We assume that the p-dimensional observations $\mbf{Y}^{(k)} (k=1,\cdots, n)$ can be modeled
as:
\begin{equation}\label{eq:GAR_1_norm}
	\mbf{Y}^{(k)} = (\theta_0 \mathbb{I}_p + \theta_1 \mathbb{L})^{-1} \mbf{Z}^{(k)}, ~~~ \theta_0>0, ~\theta_1>0,   ~\mbb{L} \in \mathcal{L}_{\mathcal{N},p}
\end{equation}
where $\mbf{Z}^{(k)} (k=1,\cdots, n)$ are i.i.d. $N(\mathbf{0}_p,\mathbb{I}_p)$ and the parameters $\theta_0, \theta_1$ are referred to as the \textit{graph filter-parameters}.

Recall that $\mcl{L}_{\mathcal{N}, p}$ denotes the space of normalized Laplacian matrices associated 
with weighted, undirected graphs with $p$ nodes. By Lemma \ref{lemma: L_N_space}, 
matrices belonging to $\mcl{L}_{\mathcal{N}, p}$  are positive semi-definite (p.s.d.). 
The GAR model (\ref{eq:GAR_1_norm}) implies that the covariance of the observations satisfies
\begin{equation}
	\label{eq:GAR_1_covariance}
\Sigma = \mbox{Var}(\mbf{Y}^{(k)}) = (\theta_0 \mathbb{I}_p +\theta_1 \mathbb{L})^{-2} 
\end{equation}
and thus co-diagonalizes with a normalized graph Laplacian matrix $ \mathbb{L}$. 

The main goal in this paper is to estimate  $ \mbb{L}$  and its zero pattern based on the observations $\{\mbf{Y}^{(k)}\}_{k=1}^n$ and thus reconstruct the underlying graph $\mathbb{G}$ defined through the zero-pattern of the off-diagonal entries of $\mathbb{L}$. 

In the following, we first discuss the interpretation and identifiability of the GAR model. We then propose an estimation procedure and a goodness-of-fit measure in Section \ref{subsec:estimation_Laplacian_N}. 
\subsubsection*{Interpretation}
\begin{figure}
	\begin{center}
		\caption{  (a) stationary time series; (b) stationary graph-referenced processes 	\label{fig:AR_analogy}}
		\includegraphics[width=4in]{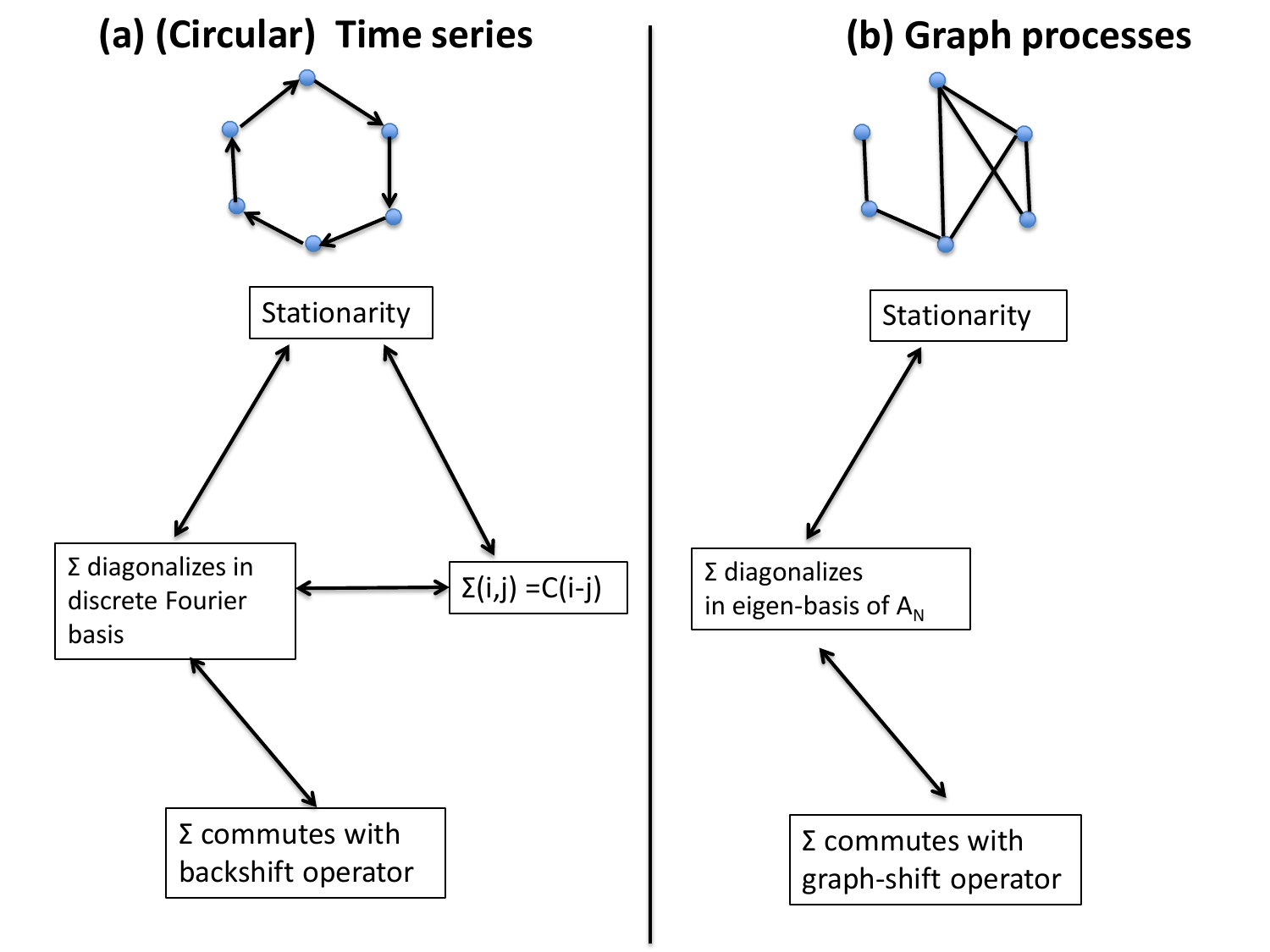} 		
	\end{center}

\end{figure}

	The GAR  model  (\ref{eq:GAR_1_norm}) is analogous  to the  AR(1) model in time series.  For $\mathbb{L} \in \mcl{L}_{\mathcal{N}, p}$, by definition, there exists a weighted adjacency matrix $\mathbb{A} \in \mathcal{A}_p$, such that $\mathbb{L}=\mathbb{I}_p - \mathbb{A}_{\mathcal{N}}$, where  $ \mathbb{A}_{\mathcal{N}}$ is the degree normalized version of $\mathbb{A}$. Therefore,  $	\mbf{Y} = (\theta_0\mbb{I}_p + \theta_1 \mbb{L})^{-1} \mbf{Z} $ can be expressed as 
		\begin{eqnarray}
			\label{eq:AR_analogy}
		\mbf{Y} 
		&= &\left((\theta_0+\theta_1)\mbb{I}_p - \theta_1 \mbb{A}_{\mathcal{N}}\right)^{-1} \mbf{Z} \nonumber \\
		&=& \sigma (\mbb{I}_p- \beta \mbb{A}_{\mcl{N}})^{-1} \mbf{Z}, ~~~\text{where}~ \sigma = 1/(\theta_0+\theta_1), ~\beta = \theta_1/(\theta_0+\theta_1) \nonumber\\
		&=& \sigma \left(\mbb{I}_p +\beta \mbb{A}_{\mcl{N}}  (\mbb{I}_p- \beta \mbb{A}_{\mcl{N}})^{-1}\right) \mbf{Z} \nonumber\\
		& =&  \beta \mbb{A}_{\mcl{N}} \sigma (\mbb{I}_p- \beta \mbb{A}_{\mcl{N}})^{-1} \mbf{Z}+\sigma \mbf{Z} \nonumber\\
		&=&\beta \mbb{A}_{\mcl{N}} \mbf{Y} + \sigma \mbf{Z}.
	\end{eqnarray}
	The third equality holds because $0<\beta <1$ and the eigenvalues of the normalized adjacency matrix $\mbb{A}_{\mcl{N}}$ are between $-1$ and $1$ \citep{Chung1997}.  
	Moreover, it is clear from equation (\ref{eq:AR_analogy}) that if $\mbb{A}_{\mcl{N}}$ is replaced by the (cyclic permutation) matrix associated with the back-shift operator,  we have an (cyclic) AR(1) process. 
	 Thus, 
	treating $\mathbf{Z}$ as an innovation term, the representation  (\ref{eq:AR_analogy}) justifies
	the nomenclature of a \textit{graphical autoregressive
		model (of order 1)}.  We also refer   to  $\mbb{A}_{\mcl{N}}$  as the \textit{graph-shift operator}.  This analogy is illustrated in Figure \ref{fig:AR_analogy}. 
Note that, not  all characterizations of stationarity under time series could be generalized to graphs. Particularly, in stationary time series,  $\Sigma(i,j)=C(i-j)$ for some symmetric nonnegative definite function $C(\cdot)$. However, this is not the case for a general graph, as there is no (global) translation operator on undirected graphs.

\subsubsection*{Identifiability}
Under the GAR model (\ref{eq:GAR_1_norm}), the parameter $\theta_0>0$ is identifiable as  $\theta_0^2=\lambda_{\min}(\Sigma^{-1})=\frac{1}{\lambda_{\max}(\Sigma)}$. However,  the parameters $\theta_1>0$ and $\mbb{L} \in \mcl{L}_{\mcl{N},p}$ are only partially identifiable. This is because, by Lemma \ref{lemma: L_N_space}, if $\mbb{L}$ is a normalized graph Laplacian matrix, then $c\cdot \mbb{L}$ for any $c>0$ such that $c \cdot \max_{i} \mbb{L}_{ii} \leq 1$ is also a normalized graph Laplacian matrix. However, if the graph does not contain any self loop, i.e., $w_{ii} \equiv 0$ for all $i=1,\cdots, p$, then we can show $\mbb{L}_{ii} \equiv 1$ and the GAR model becomes identifiable. We have the following lemma. 
\begin{lemma}
	\label{lemma:identifiability}
	In the GAR model (\ref{eq:GAR_1_norm}), the parameter $\theta_0$ is identifiable and $\theta_1 \cdot \mbb{L}$ is identifiable. 
If we further assume  the graph does not have any self loop, then all parameters in the GAR model  (\ref{eq:GAR_1_norm}) are identifiable. 
 
\end{lemma}
The proof of Lemma \ref{lemma:identifiability} is given in Appendix \ref{sec: appendix_lemma}.  Because a positive constant multiplier does not change the zero-pattern of a matrix, the topology of the latent graph (except for potential self-loops) is identifiable under the GAR model.

\subsubsection*{Reparametrization}
Since by Lemma  \ref{lemma:identifiability}, $\theta_1$ and $\mbb{L}$ are only identifiable up to positive  multipliers, in the following, we absorb $\theta_1$ into $\mbb{L}$ and consider a modified parameter space for $\mbb{L}$, i.e., the GAR model (\ref{eq:GAR_1_norm}) is reparameterized as: For $k=1,\cdots, n$
\begin{equation}
	\label{eq:GAR_norm_repara}	
	\mbf{Y}^{(k)} = (\theta_0 \mathbb{I}_p + \mathbb{L})^{-1} \mbf{Z}^{(k)}, ~~~ \theta_0>0, ~\mbb{L} \in \widetilde{\mathcal{L}}_{\mathcal{N},p}, 
\end{equation}
where  $\mbf{Z}^{(k)} (k=1,\cdots, n)$ are i.i.d. $N(\mathbf{0}_p,\mathbb{I}_p)$ and 
\begin{eqnarray}
	\widetilde{\mathcal{L}}_{\mathcal{N},p}:= \bigl\{ \mbb{L} \in \mbb{R}^{p \times p}: \mbb{L} = \mbb{L}^T; ~ \mbb{L}_{ij} \leq 0~\mbox{for all}~ 1 \leq i \neq j \leq p; \nonumber \\
	\exists ~ \mbf{v}_0 \succ \mbf{0}, s.t. ~\mbb{L} \mbf{v_0} = \mbf{0}_p \bigr\}
\end{eqnarray}

\subsection{Estimation procedure}
\label{subsec:estimation_Laplacian_N}

Since we are interested in the settings where the dimension $p$ could go to infinity with the sample size $n$, we propose to estimate $(\theta_0, \mbb{L})$ under (\ref{eq:GAR_norm_repara}) through minimizing the $\ell_1$ penalized negative log-likelihood (up to constants) :
\begin{equation}
	\label{eq: GAR_penloglike}
	g_\lambda(\theta_0,\mbb{L}) = \frac{1}{2} \tr((\theta_0 \mbb{I}_p + \mbb{L} )^{2}\widehat{\Sigma}) - \log\det(\theta_0 \mbb{I}_p +  \mbb{L}) + \lambda \|\mbb{L}\|_{1,off},
\end{equation}
where $
\|\mbb{L}\|_{1,off} := -\sum_{i\neq j} \mbb{L}_{ij} =\tr(\mbb{L}(\mbb{I}_p-\mbf{1}_p \mbf{1}_p^T))$; 
$\lambda>0$ is a tuning parameter; and $$
\widehat{\Sigma}=\frac{1}{n}\sum_{k=1}^n (\mbf{Y}^{(k)} -\overline{\mbf{Y}}) (\mbf{Y}^{(k)} -\overline{\mbf{Y}})^T, ~~~ \overline{\mbf{Y}}=\frac{1}{n} \sum_{k=1}^n \mbf{Y}^{(k)}
$$ 
are the sample covariance matrix and  the sample mean vector, respectively. 

 Minimizing (\ref{eq: GAR_penloglike}) directly on the parameter space $\mathbb{R}^{+} \times 	\widetilde{\mathcal{L}}_{\mathcal{N},p}$ is challenging as this space is non-convex due to the $\mbf{v}_0$ requirement (note that $\mbf{v}_0$ is unknown). 
In this subsection, we describe a multi-step  estimation procedure for the GAR model (\ref{eq:GAR_norm_repara})  which starts with an initial estimator of $\theta_0$, followed by a sparse  estimator of $\mbb{L}$ in a ``relaxed" parameter  space that drops the $\mbf{v}_0$ requirement, then followed by constrained estimators of $\mbb{L}$ and $\theta_0$  given the zero-pattern from the previous step. The rationale of this estimation procedure is to (i) address the challenge associated with estimating the eigenvector $\mbf{v}_0$ by separating it from $\mbb{L}$ and $\theta_0$ estimation; (ii) infer a latent graph by sparse penalized MLE; (iii) improve the estimation of $\theta_0$ and that of the non-zero entries of $\mbb{L}$ through ``post-selection" re-estimation.

\subsubsection*{Step 0: Initial estimator of $\theta_0$}
In this step, we obtain an initial estimator for $\theta_0$ through the largest eigenvalue of the sample covariance matrix $\widehat{\Sigma}$:
$$
\wh{\theta}_0^{(ini)}:=\frac{1}{\sqrt{\lambda_{\max}(\widehat{\Sigma})}}
$$

\subsubsection*{Step 1: Graph topology estimation}

In this step, we obtain a sparse estimator for  $\mbb{L}$  through an $\ell_1$ penalized MLE on a relaxed parameter space with $\theta_0$ fixed at $\hat{\theta}_0^{(ini)}$. We then perform post-estimation thresholding to enhance sparsity and obtain the estimated graph. The purpose of this step is to estimate the topology of the underlying graph defined through the zero-pattern of the off-diagonal entries of $\mbb{L}$. 

Specifically, define the\textit{ relaxed parameter space} by dropping the $\mbf{v}_0$ requirement: 
\begin{equation}
	\label{eq: GAR_step1_space}
			\widetilde{\mcl{L}}^{+}_{\mathcal{N}, p} := \bigl\{  \mbb{L} \in \mbb{R}^{p \times p}: \mbb{L} ~ \text{p.s.d.}; ~\mbb{L} = \mbb{L}^T; ~ \mbb{L}_{ij} \leq 0~\mbox{for all}~ 1 \leq i \neq j \leq p  \bigr\}
\end{equation}
Note that, $	\widetilde{\mcl{L}}^{+}_{\mathcal{N}, p}$ is a convex subset of  the original parameter space	$\widetilde{\mathcal{L}}_{\mathcal{N},p}$. 
The Step 1 estimator for $\mbb{L}$ is defined as 
\begin{equation}
\label{eq:GAR_step1_estimator}
\wh{\mbb{L}}_\lambda:=\text{argmin}_{\mbb{L} \in \widetilde{\mcl{L}}^{+}_{\mathcal{N}, p}} g_\lambda(\hat{\theta}_0^{(ini)},\mbb{L}),
\end{equation}
which is a convex optimization problem with respect to $\mbb{L}$ and can be solved by an ADMM Algorithm \ref{alg:ADMM_L2} \citep{BoydEtAl2011}. See Appendix \ref{subsec:ADMM_sep} for details. 

We then define the ``null edge" set for the latent graph as:
\begin{equation}
\label{eq:null_set}
	\mathcal{N}_{\lambda, \epsilon_{thre}}:=\{(i,j): 1 \leq i \not = j \leq p~s.t.~ |\wh{\mbb{L}}_{\lambda, ij}| \leq \epsilon_{thre}\},
\end{equation} 
where $\epsilon_{thre}>0$ is a thresholding parameter used to  set small noisy estimates to zero and consequently  improve graph topology estimation.  

Finally, the estimated graph,  $
\wh{\mbb{G}}_{\lambda, \epsilon_{thre}}
$,  over the node set $\mbb{V}=\{1,\cdots, p\}$, is defined as having an edge $i \sim j$ if and only if $(i,j) \notin 	\mathcal{N}_{\lambda, \epsilon_{thre}}$.

\subsubsection*{Step 2:  Constrained estimator of $\mbb{L}$}
Given the ``null edge" set from Step 1, we then update the estimator of $\mbb{L}$ on a \textit{constrained parameter space}:
\begin{eqnarray}
	\widetilde{\mcl{L}}^{c,+}_{\mathcal{N}, p} := \bigl\{  \mbb{L} \in \mbb{R}^{p \times p}: \mbb{L} ~ \text{p.s.d.}; ~\mbb{L} = \mbb{L}^T; ~ \mbb{L}_{ij} \leq 0~\mbox{for all}~ 1 \leq i \neq j \leq p \nonumber \\
	\mbb{L}_{ij}=0, ~\mbox{for}~ (i,j) \in \mathcal{N}_{\lambda, \epsilon_{thre}}
	 \bigr\},
\end{eqnarray}
which is a convex subset of  the original parameter space	$\widetilde{\mathcal{L}}_{\mathcal{N},p}$. 

Specifically, the updated Step 2 estimator for $\mbb{L}$ is 
	 \begin{equation}
	 	\label{eq:GAR_step2_estimator}
\wt{\mbb{L}}_{\lambda, \epsilon_{thre}}=\text{argmin}_{\mbb{L} \in \wt{\mcl{L}}^{c,+}_{\mathcal{N}, p}} g(\hat{\theta}_0^{(ini)},\mbb{L}),
\end{equation}
where $g(\cdot, \cdot )$ is the (unpenalized) negative loglikelihood (up to constants) under the GAR model: 
\begin{equation}
	\label{eq: GAR_loglike}
	g(\theta_0,\mbb{L}) = \frac{1}{2} \tr((\theta_0 \mbb{I}_p + \mbb{L} )^{2}\widehat{\Sigma}) - \log\det(\theta_0 \mbb{I}_p +  \mbb{L}).
\end{equation}
 
 \subsubsection*{Step 3: Joint estimation of $(\theta_0, \mbb{L})$}
In this step, we first obtain an estimator $\widehat{\mbf{v}}_{\lambda, \epsilon_{thre}}$ of the $\mbf{v}_0$ vector by
$$
\label{eq: GAR_step3_estimator_v0}
\text{minimize} \parallel \wt{\mbb{L}}_{\lambda, \epsilon_{thre}} \mbf{v}_0 \parallel_2~~ \text{subject to}~~  \mbf{v}_0 \succ \mbf{0} ~ \text{and}~ \parallel \mbf{v}_0 \parallel_2 =1, 
$$
where $\parallel \cdot \parallel_2 $ denotes the $\ell_2$ norm and $\wt{\mbb{L}}_{\lambda, \epsilon_{thre}}$ is the Step 2 estimator of $\mbb{L}$ (see Appendix \ref{subsec:ADMM_v0} for Algorithm \ref{alg:ADMM_deg}). 

We then obtain a \textit{constrained ``estimated" parameter space}:  
	\begin{eqnarray}
	\widehat{\mcl{L}}^{c}_{\mathcal{N}, p} := \bigl\{ \mbb{L} \in \mbb{R}^{p \times p}: \mbb{L} ~ \text{p.s.d.}; ~\mbb{L} = \mbb{L}^T; ~ \mbb{L}_{ij} \leq 0~\mbox{for all}~ 1 \leq i \neq j \leq p \nonumber \\ 
	\mbb{L}_{ij}=0, ~\mbox{for}~ (i,j) \in \mathcal{N}_{\lambda, \epsilon_{thre}}; 
		\mbb{L} \widehat{\mbf{v}}_{\lambda, \epsilon_{thre}}=\mbf{0}_p  \bigr\},
\end{eqnarray}
which is a convex subset  of  the original parameter space	$\widetilde{\mathcal{L}}_{\mathcal{N},p}$.

Finally, the Step 3 estimator for $(\theta_0, \mbb{L})$ is defined as: 
\begin{equation}
            \label{eq: GAR_step3_estimator_v0_known}
			(\wh{\mbb{L}}_{\lambda, \epsilon_{thre}},\wh{\theta}_{0, \lambda, \epsilon_{thre}})=\text{argmin}_{\mbb{L} \in \widehat{\mcl{L}}^c_{\mathcal{N}, p}, \theta_0>0} g(\theta_0,\mbb{L}). 
\end{equation}
This minimization is bi-convex with respect to $\theta_0$ and $\mbb{L}$ and can be solved by an ADMM Algorithm \ref{alg:ADMM_Lap} (see Appendix \ref{subsec:ADMM_simu}).

\subsubsection*{Tuning parameters selection}
The proposed estimation procedure involves two tuning parameters, $\lambda>0$ in the $\ell_1$ penalty when deriving the Step 1 estimator $\widehat{\mbb{L}}_{\lambda}$,  and $\epsilon_{thre}>0$ for thresholding  the off-diagonal entries of $\widehat{\mbb{L}}_{\lambda}$  to derive the estimated graph  $
\wh{\mbb{G}}_{\lambda, \epsilon_{thre}}$. We propose to use the \textit{extended BIC (eBIC)} (Chen, 2011) to choose these two tuning parameters. Specifically, for the Step 3 estimator 	$(\wh{\mbb{L}}_{\lambda, \epsilon_{thre}},\wh{\theta}_{0, \lambda, \epsilon_{thre}})$, the corresponding eBIC score is defined as 
\begin{equation}
	\label{eq:GAR_eBIC}
	\text{eBIC}(\lambda, \epsilon_{thre}) :=2n \cdot g(\wh{\theta}_{0, \lambda, \epsilon_{thre}}, \wh{\mbb{L}}_{\lambda, \epsilon_{thre}})+\hat{s}\cdot \log(n)+2 \gamma \cdot \log \frac{Q!}{(Q-\hat{s})! \hat{s}!},
	\end{equation}
where $0 \leq  \gamma \leq 1$ is a constant,	$Q =\frac{p(p-1)}{2}$ is the total number of potential edges in the graph, and $\hat{s} := |\wh{\mbb{G}}_{\lambda, \epsilon_{thre}}|$ is the number of edges in the estimated graph $\wh{\mbb{G}}_{\lambda, \epsilon_{thre}}$.

Note that when $p \gg \hat{s}$, the eBIC criterion in (\ref{eq:GAR_eBIC}) can be approximated by 
$$
2n \cdot g(\wh{\theta}_{0, \lambda, \epsilon_{thre}}, \wh{\mbb{L}}_{\lambda, \epsilon_{thre}})+\hat{s}\cdot \log(n)+4 \gamma \hat{s} \log p.
$$
This last expression closely resembles the eBIC criterion proposed by \cite{FoygelD2010} for model selection of GGM. They showed  that, with the appropriate choice of $\gamma$, using eBIC criteria leads to consistent model selection.

In our context, we set $\gamma=0.5$ if $p/n \leq 0.5$ (lower-dimensional setting) and $\gamma=1$ if $p/n>0.5$ (higher-dimensional setting). We choose the pair of tuning parameters that minimizes (\ref{eq:GAR_eBIC}), denoted by $(\lambda^{\ast}, \epsilon_{thre}^{\ast})$. The final estimators are  $\wh{\mbb{L}}:=\wh{\mbb{L}}_{\lambda^{\ast}, \epsilon_{thre}^{\ast}},~\wh{\theta}_0:=\wh{\theta}_{0, \lambda^{\ast}, \epsilon_{thre}^{\ast}}$.

\subsubsection*{Goodness-of-fit measure}

Here we consider a goodnes-of-fit measure for the GAR model through parametric bootstrap. Given $\lambda$ (say $\sqrt{\log p/n}$),  we derive the Step 1 estimator $\wh{\mbb{L}}_\lambda$ and the corresponding plugin estimator for $\Sigma$, denoted by $\wh{\Sigma}_\lambda$. We then calculate the loglikelihood of the observed data with respect to $N(\mathbf{0}, \wh{\Sigma}_\lambda)$  and denote it by $\ell^{obs}$.  
We next draw $B$  (e.g., $B=100$)  bootstrap samples of size $n$  from $N(\mathbf{0}, \wh{\Sigma}_\lambda)$. For the $b$th bootstrap sample ($b=1,\cdots, B$), we derive the  Step 1 estimator for $\mbb{L}$ and the plugin estimator for $\Sigma$, denoted by $\wh{\Sigma}^{(b)}_{\lambda}$. We then calculate the loglikelihood of the $b$th bootstrap sample with respect to $N(\mathbf{0}, \wh{\Sigma}^{(b)}_{\lambda})$, denoted by  $\ell^{(b)}$. 
Finally, we calculate a goodness of fit measure: 
\begin{equation}
\label{eq:goodness-of-fit}
GF=\frac{\sum_{b=1}^B \text{I}(\ell^{(b)} \leq \ell^{obs})}{B},
\end{equation}
where $\text{I}(\cdot)$ is the indicator function. 
If GF is close to $1$, then it indicates that the GAR model is a good fit for the given data, whereas if $GF$ is close to $0$, then it indicates that the GAR model does not fit the data well and is likely mis-specified.

Through simulation experiments, we observe that the GF measure is effective when $p \leq n$; however, this measure fails when $p > n$ (Table \ref{tab:fit_parbootstrap}). 

As an alternative for $p > n$, we propose fitting a GGM and using the eBIC score to determine whether the GAR model should be used.  As shown in Table  \ref{tab:GAR_vs_glasso_ebic_results}, when the data are generated according to a GAR model, the eBIC score of the selected GAR model is consistently smaller than that of the selected GGM. On the other hand, when the data are generated from a power-law GGM (i.e., not a GAR model), the eBIC score for the GAR model is consistently larger than that for the GGM.

\section{GAR Estimation Consistency}\label{sec:asymp_Laplacian}
\label{sec:consitency}

In the following, we discuss the consistency of the Step 1 estimator $\wh{\mbb{L}}_{\lambda}$ (\ref{eq:GAR_step1_estimator}) under the GAR model (\ref{eq:GAR_norm_repara}). Throughout we use $\theta_0^\ast$ and $\mbm{L}^{\ast}$ to indicate the true values of $\theta_0$  and $\mbm{L}$, respectively.   Let $\mcl{S} = \{(i,j): 1\leq i \neq j \leq p ~\mbox{such that}~ \mbb{L}^{\ast}_{ ij} < 0\}$ denote the set of non-zero off-diagonals of $\mbm{L}^{\ast}$,
	and let $s = |\mcl{S}|$ denote the cardinality of $\mcl{S}$.

We make the following assumptions: 
\begin{itemize}
\item[\tb{A1}]  $\| \mbm{L}^\ast \| \leq K_1$, for some $K_1 \in (0,\infty)$ not depending on $n,p,s$.
\item[\tb{A2}] $ 0< K_2/2  \leq \theta_0{^\ast} \leq K_2$ for some  $K_2 \in (0,\infty)$ not depending on $n,p,s$.
\item[\tb{A3}] $s \log p = o(n)$ as $p,n \to \infty$.
\end{itemize}
Assumptions \tb{A1}--\tb{A2} imply that the true covariance $\Sigma^{\ast}=(\theta_0^{\ast}\mbb{I}_p+ \mbb{L}^{\ast})^{-2}$ has a bounded condition number. 

In the following, we present the results. The detailed proofs are deferred to Appendix \ref{sec: appendix_proof}.

\subsection{$\theta_0$ known}

First, we consider the case where the true value of $\theta_0$ is known. i.e., 
\begin{equation}
\label{eq:GAR_step1_estimator_theta0_known}
\wh{\mbb{L}}_\lambda:=\text{argmin}_{\mbb{L} \in \widetilde{\mcl{L}}^{+}_{\mathcal{N}, p}} g_\lambda(\theta_0^{\ast},\mbb{L}),
\end{equation}

We primarily follow the approach of \cite{RothmanBLZ2008} and have the following result.

\begin{theorem}\label{thm:consistency_hat_L_theta_0_known}
Suppose that \tb{A1}--\tb{A3} hold. Also let $\lambda = \lambda_n = C \sqrt{\frac{\log p}{n}}$, for an appropriate $C > 0$ (not depending on $n$, $p$ or $s$). Then, with probability tending to 1,
\begin{equation}
\| \wh{\mbm{L}}_{\lambda_n} - \mbm{L}^{\ast}\|_F = O\left(\sqrt{\frac{(s+p)\log p}{n}}\right),
\end{equation}	
\end{theorem}
where $\wh{\mbm{L}}_{\lambda_n} $ is as defined in (\ref{eq:GAR_step1_estimator_theta0_known}) with $\lambda=\lambda_n$ and $\|\cdot \|_F$ denotes the Frobenius norm.

\subsection{Plug-in $\wh{\theta}_0$}

We next consider the case when a plug-in estimator of $\theta_0$ is used in place of 
$\theta_0$ in the objective function (\ref{eq: GAR_penloglike})
and then minimization over $\widetilde{\mcl{L}}^{+}_{\mathcal{N}, p} $ (\ref{eq: GAR_step1_space}) is carried out 
to obtain an estimator of $\mbm{L}$.

Let $\wh{\theta}_0$ denote a (arbitrary) consistent estimate of $\theta_0$ with the requirement that $\wh{\theta}_0 > 0$. We make
the following additional assumption.  
\begin{itemize}
\item[\tb{B1}] There exists a sequence $\gamma_n\to 0$ such that
\begin{equation}
|\wh{\theta}_0-\theta_0^{\ast}| = O(\gamma_n) ~~\mbox{with probability}~ \to 1.
\end{equation}
\end{itemize}

Let
\begin{equation}
\label{eq:GAR_step1_estimator_theta0_plugin}
\wh{\mbb{L}}_\lambda:=\text{argmin}_{\mbb{L} \in \widetilde{\mcl{L}}^{+}_{\mathcal{N}, p}} g_\lambda(\wh{\theta}_0,\mbb{L}).
\end{equation}
We have the following result.

\begin{theorem}\label{thm:consistency_hat_L_theta_plugin}
	Suppose that \tb{A1}--\tb{A3} and \tb{B1} hold. Also let
\begin{equation}\label{eq:lambda_r_plugin_setup}
\lambda_n = C_4 \max\left\{\sqrt{\frac{\log p}{n}} , \gamma_n\right\}
\end{equation}
and $\sqrt{s+p}\lambda_n \to 0$ as $n\to \infty$. Then with probability tending to $1$,
$$
\|\wh{\mbm{L}}_{\lambda_n} - \mbm{L}^{\ast}\|_F = O(\sqrt{s+p}\lambda_n),
$$
where $\wh{\mbm{L}}_{\lambda_n} $ is as defined in (\ref{eq:GAR_step1_estimator_theta0_plugin}) with $\lambda=\lambda_n$. 
\end{theorem}

\begin{remark}
If $p = o(n)$ and $\wh{\theta}_0 = \wh{\theta}_0^{(ini)} := 1/\sqrt{\lambda_{max}(\wh{\Sigma})}$,  then by standard results on the behavior of largest eigenvalue of a large dimensional sample covariance matrix (cf. \cite{Vershynin2010}), we can find a universal bound on the difference between the largest eigenvalue of $\Sigma^*$ and that of $\wh{\Sigma}$, from which we can show that
condition \tb{B1} holds for $\gamma_n = \sqrt{\frac{\log p}{n}} + \sqrt{\frac{p}{n}}$. Note that, this $\gamma_n$ leads to a $\lambda_n$ that is larger than the optimal rate, namely $\sqrt{\frac{\log p}{n}}$, for the $\theta_0$ known case. 

However, if we assume further that
the largest eigenvalue of $\Sigma^*$ has a fixed multiplicity, and there is an asymptotically non-vanishing gap between the largest and second largest eigenvalues of $\Sigma^*$ (equivalently, a non-vanishing gap between the smallest and the second smallest eigenvalues of $\mbm{L}^*$), then
\tb{B1} holds for $ \wh{\theta}_0^{(ini)}$ by setting $\gamma_n = \sqrt{\frac{\log p}{n}} + \frac{p}{n}$. We state this result in the form of Lemma \ref{lem:eigen_rate} below. In particular, if $p^2 = O(n)$, then under this setting, $\|\wh{\mbm{L}}_{\lambda_n}- \mbm{L}^*\|_F$ can achieve the same rate  as in the case when $\theta_0$ is known. 

\end{remark}

\begin{lemma}\label{lem:eigen_rate}
Suppose conditions \tb{A1}--\tb{A3} hold. Suppose further that there is a nonvanishing gap between the smallest distinct eigenvalue 0 and the next smallest eigenvalue of $\mbm{L}^*$, and the multiplicity of the eigenvalue 0 is fixed, even as dimension $p$ increases.
Then with probability $\to 1$, 
\[
|\lambda_{max}(\wh{\Sigma}) - \lambda_{max}(\Sigma^*)| = O\left(\sqrt{\frac{\log p}{n}} + \frac{p}{n}\right),
\]
where $\lambda_{max}$ denotes the largest eigenvalue.
\end{lemma}

\section{Simulation}
\label{sec:simulation}

In this section, we conduct simulation experiments to evaluate the finite sample performance of the proposed GAR fitting procedure. We also compare it with the popular Gaussian graphical model (GGM) method, \textit{glasso} \citep{friedman2008sparse}, in terms of estimating covariance and concentration matrices, as well as detecting nonzero entries in the concentration matrix.

We present detailed results for a "baseline setting" in this section. Results for additional settings (e.g., denser graphs or graphs with self-loops) are provided in Appendix \ref{sec:appendix_simulation}. 

\subsection{Data generation}
\label{subsec:simu_data_generation}

First, we generate a random undirected graph with edge probability $2/p$ and no self-loops (i.e., $w_{ii} \equiv 0$), where $p$ is the dimension of the signal. We consider three dimensions: $p=100, 250, 500$, corresponding to graphs with $105, 307, 555$ edges, respectively. The nonzero edge weights are uniformly generated from the interval $[0.5, 1]$. We also evaluate three sample sizes: $n=100, 250, 500$. For each combination of $(n,p)$, we generate 100 independent replicates from the corresponding GAR model with $\theta_0=1$ and $\theta_1=2$.

For each replicate, we fit the GAR model using a series of tuning parameters $(\lambda, \epsilon_{thre})$, and select the final model using the eBIC criterion (\ref{eq:GAR_eBIC}) as described in Section \ref{subsec:estimation_Laplacian_N}.  Specifically, both $\lambda$ and $\epsilon_{thre}$ are of the form $C\sqrt{\frac{\log p}{n}}$. For $\lambda$, we use $C=1, 0.5$. For $\epsilon_{thre}$, we use 10 equally spaced values on a logarithmic scale, with $C$ ranging from $(0, C^{\ast})$. For $p=100, 250, 500$, the values of $C^{\ast}$ are 0.05, 0.075, and 0.1, respectively.

\subsection{Evaluation metrics}
\label{subsec:simu_eval}
For graph inference (i.e., the detection of nonzero off-diagonal entries of $\mbb{L}$), we report the \textit{false discovery rate (FDR)}, the \textit{Power}, and the  \textit{$F1$-score} defined as 
$$F1=2 \times \frac{Precision\times Recall}{Precision+Recall},$$ 
where \textit{Precision} is given by $1 - \text{FDR}$, and \textit{Recall} is equal to \textit{Power}.

We also report the estimation errors for $\theta_0, \mathbf{v}_0,$ and $\mbb{L}$ defined as 
$|\widehat{\theta}_0 - \theta^{\ast}_0|^2$, $\frac{\|\widehat{\mbb{L}} - \mbb{L}^{\ast}\|_F^2}{\|\mbb{L}^{\ast}\|_F^2}$, and  $\|\widehat{\mathbf{v}}_0 -\mathbf{v}^{\ast}_0\|^2_2$, respectively.  Note that a ``hat" on a parameter indicates the estimated value, and a ``$\ast$" denotes the true value. Furthermore, the parameter $\theta_1$ is absorbed into $\mbb{L}$ according to the reparametrized model in equation (\ref{eq:GAR_norm_repara}).

\subsection{Results}	
For the ``baseline setting", all GAR fittings converged, and the results reported below are averaged across 100 replicates. Table \ref{table: baseline_results} presents the estimation errors and graph inference results for the eBIC-selected GAR model.

In Table \ref{tab:baseline_oracle_step1}, we report the estimation errors for: (i) the initial estimator $\wh{\theta}_0^{(ini)}$ of $\theta_0$, and the Step 1 estimator (\ref{eq:GAR_step1_estimator})  of $\mbb{L}$, and (ii) the ``oracle estimator" of $\mbb{L}$, where $\theta_0$, $\mathbf{v}_0$, and the nonzero pattern of $\mbb{L}$ are set to the true values.

As shown in Table \ref{table: baseline_results}, the proposed GAR fitting procedure yields satisfactory results in both parameter estimation and graph inference, even when the dimension $p$ is several times larger than the sample size $n$. Compared to the Step 1 estimator, the proposed multi-step procedure significantly improves estimation accuracy, particularly for $\theta_0$. Furthermore, compared to the ``oracle estimator," the proposed estimator (with eBIC for model selection) achieves similar error magnitudes for $\mbb{L}$ estimation, especially when the sample size is $n = 250$ or $500$.

\begin{table}[H]
	\centering
	\begin{tabular}{|c|c|c|c|c|c|c|c|}
		\hline
		 \multicolumn{2}{|c|}{\bf Setting} & $\theta_0$  & $\mathbf{v}_0$  &  $\mbb{L}$   & {\bf Power} & {\bf FDR} & {\bf F1} \\
		 \multicolumn{2}{|c|}{}& Error &  Error  &Error & & &\\
		\hline
		 \multirow{3}{*}{$p=100$} & $n=100$ & .013 &  .048 & .020 & .946 & .050 & .948\\
		 & $n=250$ & $<.001$ & .016 & .006 & .997 & .036 & .980\\ 
		  & $n=500$ & $<.001$ & .010 & .003  & 1 & .013  & .993\\
  \cline{1-8}
		 \multirow{3}{*}{$p=250$} & $n=100$ & .036 & .066 &  .034 &.880 & .063 & .908\\
		& $n=250$ &  .005 &  .023 & .008 & .984 & .013 & .985\\ 
		  & $n=500$ & $<.001$ & .015 &  $.003$ & $>.999$ & .024 & .988\\ 
  \cline{1-8}
		 \multirow{3}{*}{$p=500$} & $n=100$ &.013 & .078 & .030 & .910 &.159  & .874\\
		 & $n=250$ & .014 & .034 & .008 & .983 & .031 & .976\\ 
		  & $n=500$ & $.008$ & .017 &  .003 & .998 & .010 & .994 \\ 
		\hline

	\end{tabular}
	\caption{Baseline simulation: Results for the eBIC-selected GAR model. 
 \label{table: baseline_results}}

\begin{table}[H]
    \centering
    \begin{tabular}{|c|c|c|c|c|}
        \hline
        \multicolumn{2}{|c|}{\bf Setting} & $\wh{\theta}_0^{(ini)}$ & Step 1 $\wh{\mbb{L}}_{\lambda}$ & {\bf Oracle} $\wh{\mbb{L}}$\\ 
        \multicolumn{2}{|c|}{}  & Error & Error & Error \\ 
        \hline
       \multirow{3}{*}{$p=100$}  & $n=100$ & .0656 & .0663 & .0109 \\
        & $n=250$ & .0274 & .0297 & .0043 \\ 
        & $n=500$ & .0140 & .0157 & .0023 \\ 
        \hline
        \multirow{3}{*}{$p=250$} & $n=100$ & .1073 & .0848 & .0126 \\ 
        & $n=250$ & .0445 & .0379 & .0052 \\ 
        & $n=500$ & .0200 & .0204 & .0028 \\
        \hline
        \multirow{3}{*}{$p=500$} & $n=100$ & .2278 & .0830 & .0125 \\ 
        & $n=250$ & .1228 & .0369 & .0053 \\ 
        & $n=500$ & .0695 & .0194 & .0027 \\
        \hline
    \end{tabular}
    \caption{Baseline simulation: Results for the Step 1 GAR estimator  ($\lambda=\sqrt{\log p/n}$) and  the oracle GAR estimator.}
    \label{tab:baseline_oracle_step1}
\end{table}

\end{table}

\subsection{Comparison with \textit{glasso}}

For $p=100, 250, 500$, the true GGM graph (defined by the nonzero off-diagonal entries of the true concentration matrix $\Omega^{\ast} = \Sigma^{\ast, -1}$) has $292, 996, 1659$ edges, respectively. For the GAR method, we obtained the plug-in estimators $\widehat{\Omega}=(\hat{\theta}_0 \mbb{I}_p + \widehat{\mbb{L}})^2$ and  $\widehat{\Sigma}=(\hat{\theta}_0 \mbb{I}_p + \widehat{\mbb{L}})^{-2}$. Moreover,  the estimated GGM graph is defined by the nonzero off-diagonal entries of $\widehat{\Omega}$.

We fitted the \textit{glasso} model under a series of tuning parameters (that cover the true model).  We then refitted the non-zero entries of the estimated concentration matrix and used eBIC to choose the tuning parameter.   
We considered the following error metrics  $\frac{\|\hat{\Omega} - \Omega^{\ast}\|_F^2}{\|\Omega^{\ast}\|_F^2}$ and 
	$\frac{\|\hat{\Sigma} - \Sigma^{\ast}\|_F^2}{\|\Sigma^{\ast}\|_F^2}$ for $\Omega$ and $\Sigma$ estimation, respectively. We also report FDR, Power and the $F1$-score for GGM graph inference.  
	
	As can be seen from Table \ref{tab:GAR_v_glasso}, the GAR estimator performs satisfactorily for both covariance matrix and concentration matrix estimation, as well as for GGM graph inference. The GAR results are also (much) better than the \textit{glasso} results, especially in terms of GGM graph inference and covariance matrix estimation. This is because the GAR estimator takes advantage of the true underlying model, which is sparser than the corresponding GGM. Additionally, the selected GAR models have smaller eBIC scores than the selected \textit{glasso} models  (Table \ref{tab:GAR_vs_glasso_ebic_results}).

\begin{table}[h]
    \centering
    \begin{tabular}{|c|c|c|c|c|c|c|c|}
    \hline
    \multicolumn{2}{|c|}{\bf Setting} & \textbf{Fitted} & $\Sigma$ & $\Omega$ & \textbf{FDR} & \textbf{Power}  & \textbf{F1} \\ 
    \multicolumn{2}{|c|}{} &  \textbf{Model} & Error & Error & &&\\ 
    \hline

    \multirow{6}{*}{$p=100$} & \multirow{2}{*}{$n=100$} &  GAR & .046 & .035 & .074 & .869 & .896 \\
    && glasso & .146 & .054 & .143 & .334 & .481 \\ 
    \cline{2-8}
    & \multirow{2}{*}{$n=250$} & GAR & .009 & .011 & .052 & .992 & .969 \\
    && glasso & .048 & .028 & .049 & .377 & .540 \\ 
    \cline{2-8}
    & \multirow{2}{*}{$n=500$} & GAR & .005 & .005 & .020 & 1 & .990 \\ 
    && glasso & .031 & .024 & .013 & .377 & .546 \\ 
    \hline
    \multirow{6}{*}{$p=250$} & \multirow{2}{*}{$n=100$} & GAR & .080 & .049 & .101 & .741 & .812 \\ 
    && glasso & .237 & .076 & .170 & .251 & .385 \\ 
    \cline{2-8}
    & \multirow{2}{*}{$n=250$} & GAR & .017 & .013 & .020 & .956 & .968 \\
    && glasso & .078 & .034 & .061 & .310 & .466 \\ 
    \cline{2-8}
    & \multirow{2}{*}{$n=500$} & GAR & .007 & .006 & .037 & .999 & .981 \\ 
    && glasso & .053 & .027 & .026 & .319 & .481 \\ 
    \hline
    \multirow{6}{*}{$p=500$} & \multirow{2}{*}{$n=100$} & GAR & .077 & .053 & .266 & .781 & .757 \\ 
    && glasso & .316 & .087 & .292 & .245 & .364 \\ 
    \cline{2-8}
    & \multirow{2}{*}{$n=250$} & GAR & .034 & .013 & .050 & .945 & .948 \\ 
    && glasso & .087 & .034 & .137 & .329 & .476 \\ 
    \cline{2-8}
    & \multirow{2}{*}{$n=500$} & GAR & .020 & .006 & .015 & .992 & .989 \\ 
    && glasso & .038 & .025 & .051 & .349 & .511 \\ 
    \hline
    \end{tabular}
    \caption{Baseline simulation: Results by GAR estimator and by \textit{glasso} estimator.}
    \label{tab:GAR_v_glasso}
\end{table}

\newpage

\section{Real Data Application}

\label{sec:application}

\begin{figure}[h]
		\centering
		\includegraphics[width=4in]{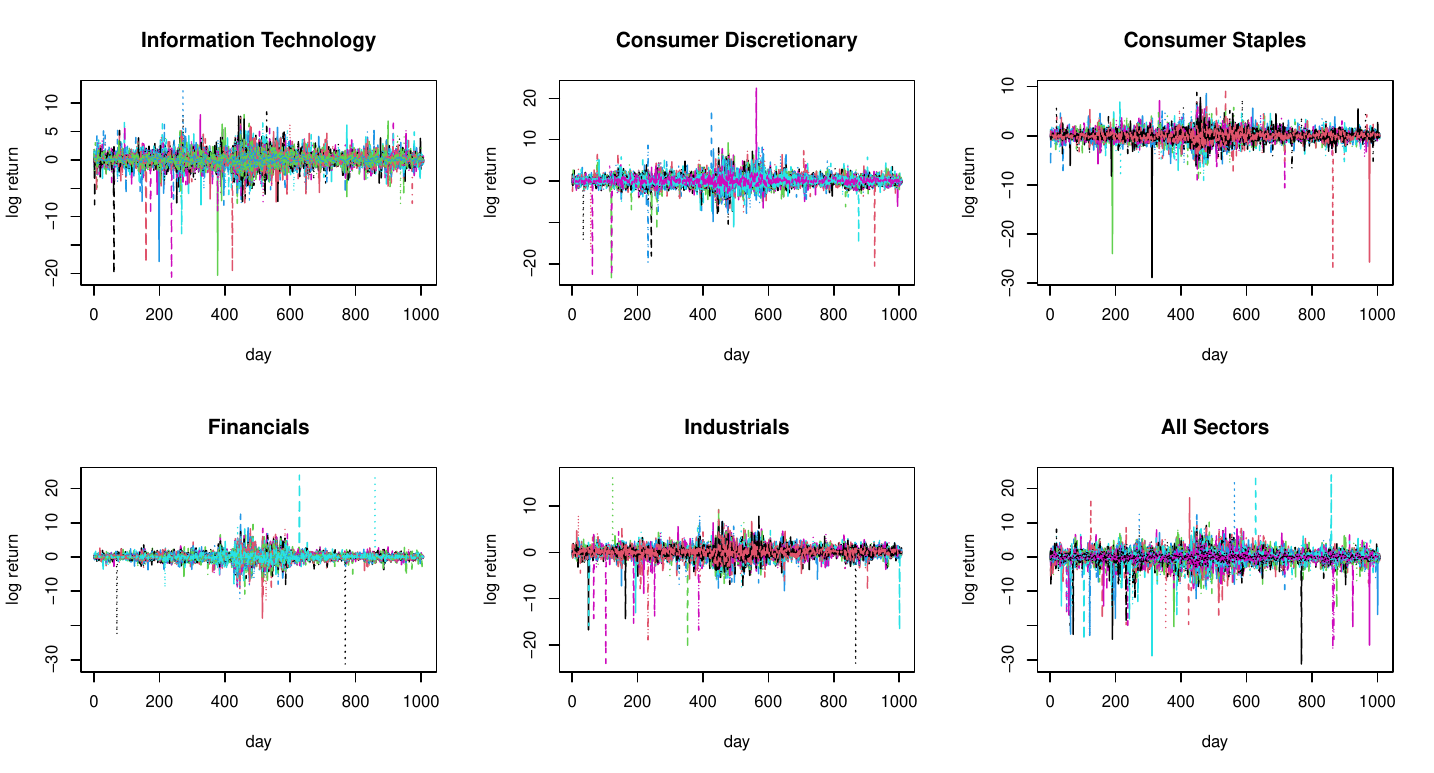}
  	\caption{S\&P 500 application. Log-return of $283$ S\&P 500 stocks from 1/1/2007 to 12/31/2010 \label{fig: sp500_logreturn}}
	\end{figure}

In this section, we fit the GAR model to an S\&P 500 stock price dataset, obtained via the R package ``quantmod" from Yahoo Finance \citep{yang2020estimating}. Specifically, We focus on 283 stocks across five GICS sectors: 58 from Information Technology, 72 from Consumer Discretionary, 32 from Consumer Staples, 59 from Financials, and 62 from Industrials. The dataset spans January 1, 2007, to January 1, 2011, covering the global financial crisis, with 1007 closing prices per stock. For the subsequent analysis, we use the logarithm of the price ratio between consecutive trading days (Figure \ref{fig: sp500_logreturn}). By examining the autocorrelations, we conclude that the independence assumption is reasonably satisfied.  Moreover, each stock's log-return is standardized to have mean zero and variance one. 

\begin{figure}[H]
		\centering
		
		\includegraphics[width=3in]{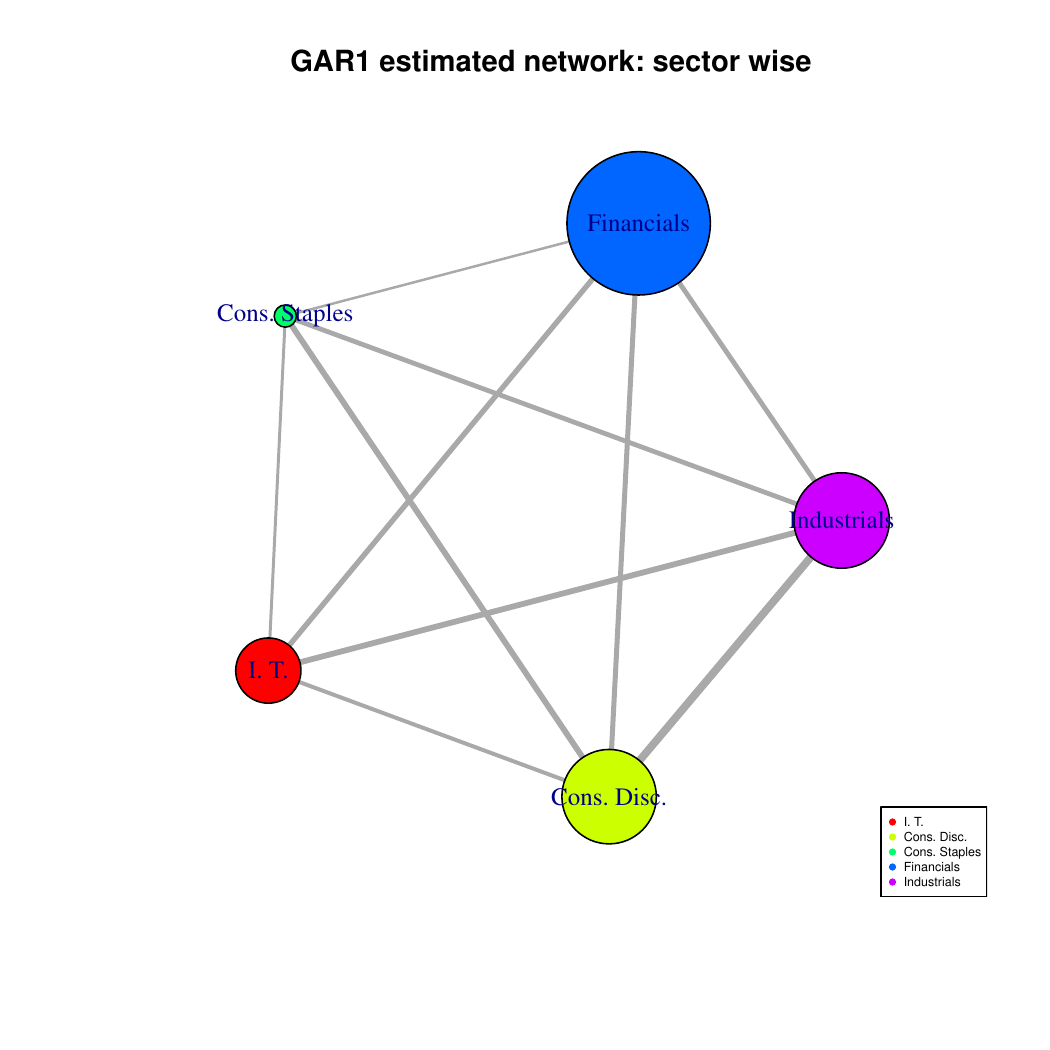}
		\caption{S\&P 500 stock price data application: GAR estimated network. Node size indicates within-sector connectivity level and edge width indicates between-sector connectivity level \label{fig: sp500_gar}}
	\end{figure}

Figure \ref{fig: sp500_gar} shows the within- and cross- sector connectivity among the 5 GICS sectors by the GAR estimated graph. In the figure, each node represents a sector and the node size is proportional to the percentage of presence of within-sector edges (defined as
the detected number of edges between two stocks in this section divided by the total number of possible
edges within this section), whereas the edge width
is proportional to the percentage of presence of cross-sector edges (defined as
the detected number of edges between two sectors divided by the total number of possible
edges between these two sectors).  Table \ref{table: sp500_gar} presents these percentages in numbers.
It can be seen that, during this period, Consumer Staples had
the least amount of within-sector connectivity, whereas Financials had
the highest within-sector connectivity. As for cross-sector connectivity, the connections between Consumer Staples and Financials/I.T. had been the weakest, whereas the connection between Consumer Discretion and Industrials had been the strongest. 

		\begin{table}[h]
		\begin{center}
			\begin{tabular}{|*{6}{c|}}
				 \cline{1-6}
				        &     I. T.& Cons. Disc.& Cons. Staples &Financials& Industrials\\\hline
				I. T.   &      \textbf{12.0\%}    &     1.9\%&           1.6\%&        2.1\%&         2.5\%\\
				Cons. Disc.  &   1.9\%     &  \textbf{14.9\%}     &      3.0\%    &    2.1\%   &      3.4\%\\
				Cons. Staples&   1.6\%      &   3.0\% &       \textbf{ 4.9\%}    &    1.3\% &        2.5\%\\
				Financials  &    2.1\%      &   2.1\%     &      1.3\%   &    \textbf{17.0\%}     &    2.0\%\\
				Industrials  &   2.5 \%     &   3.4\%  &         2.5\% &       2.0\%&        \textbf{14.2\%}\\\hline
			\end{tabular}
		\end{center}	
  \caption{S\&P 500 stock price data application: GAR estimated network. Within-sector and between-sector connectivity level measured by the percentage of edges detected (relative to all possible edges) \label{table: sp500_gar} }			
	\end{table}

\begin{figure}[H]
		\centering
			
		\includegraphics[width=2.5in]{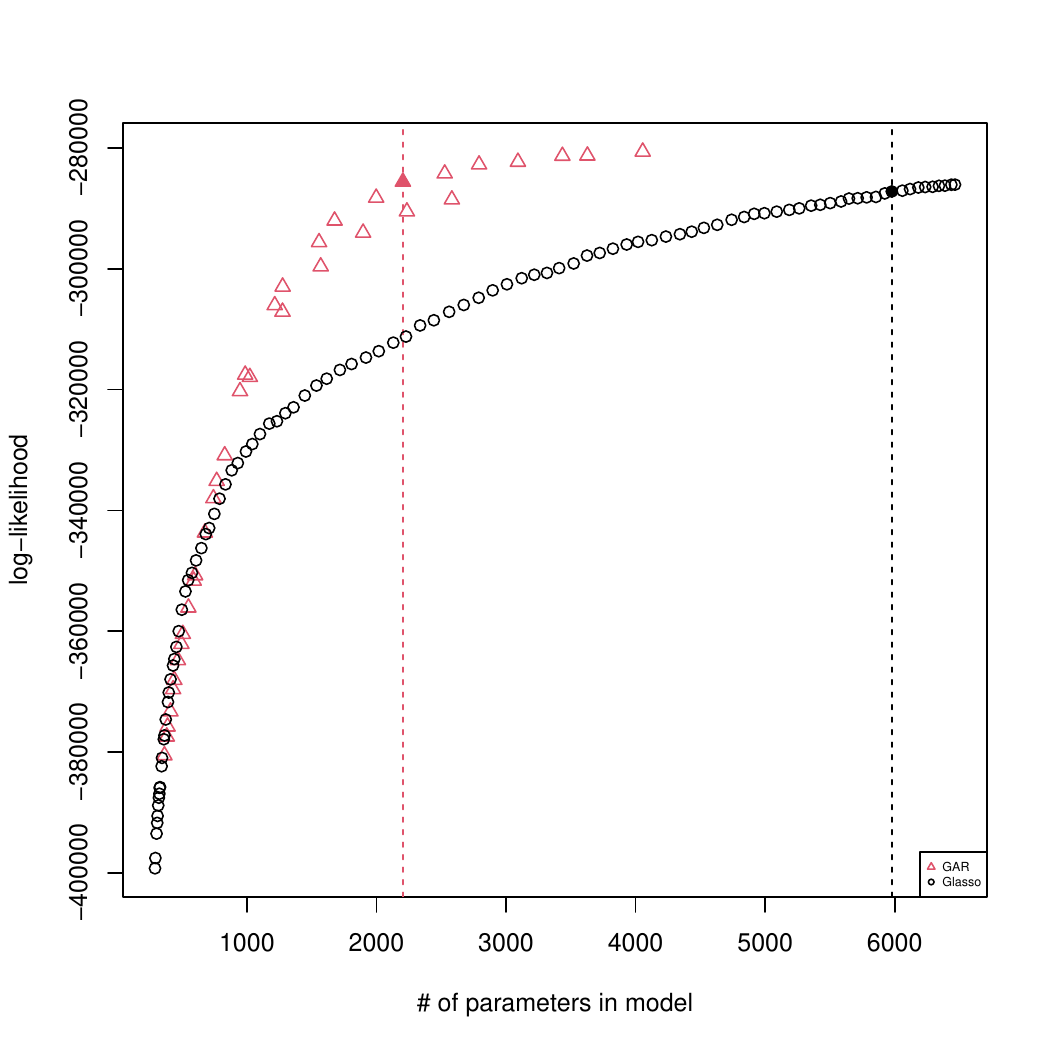}
	\caption{S\&P 500 stock price data application: log-likelihood vs.  \# of parameters. Red triangles correspond to GAR models and black circles correspond to  \textit{glasso} Models.  Dotted lines indicate eBIC selected models. \label{fig:sp500_gar_glasso}}
	\end{figure}

 Moreover, the goodness-of-fit measure (\ref{eq:goodness-of-fit}) by the parametric bootstrap procedure is equal to $1$, indicating that the GAR model fits the stock data reasonably well. We also fit \textit{glasso} to the S\&P 500 data. As can be seen from Figure \ref{fig:sp500_gar_glasso}, given the same number of parameters, the GAR model tends to have larger loglikelihood than the glasso model. Specifically, at the model selected by eBIC, the GAR model not only has much less parameters (i.e., a sparser graph) than the glasso model ($1920$ vs. $4632$), it also fitted the data better, evidenced by a larger loglikelihood ($-285605$ vs. $-290912$) and a smaller eBIC score ($592187$ vs. $622970$).

\section{Discussion}
\label{sec:discussion}

In this paper, we introduce the graphical autoregressive model (GAR) and propose a multi-step estimation procedure. We study the estimation consistency of the GAR estimator when the dimension $p$ diverges with the sample size $n$ at an appropriate rate. We apply the GAR model to S\&P 500 stock price data, which leads to a graph with higher connectivity within industrial sectors compared to between sectors. Furthermore, when compared to the widely used \textit{glasso} method, our results suggest that the GAR model offers a more efficient representation of the data when it fits well. Specifically, in the stock price application, the GAR model results in a sparser graph and provides a better fit than the graph produced by the \textit{glasso} model.

The GAR model falls under a general framework that we refer to as the \textit{Spectral Graph Models (SGM) framework}, where the graph-referenced process is modeled as
\begin{equation*}
\label{eq:sgm-framework}
\mathbf{Y}_t | \mathbb{G}_t \sim  N(\mathbf{0}_p, \Sigma_t), ~~~ \Omega_t = \Sigma^{-1}_t = g (\mathbb{L}_{t}, \boldsymbol{\theta}_t), ~~~ \mathbb{L}_{t} \in \mathcal{L}_{\mathcal{N},p}, ~~~~ t=1,\cdots, T,
\end{equation*}
where $\boldsymbol{\theta}_t$ is a finite-dimensional graph filter parameter and $g(x, \boldsymbol{\theta}_t)$ is a positive valued, strictly increasing function on   $x \in [0,2]$. Moreover, $g (\mathbb{L}_{t}, \boldsymbol{\theta}_t)$ has the same eigen-spaces as $\mathbb{L}_{t}$.  This framework says that, given the graph $\mathbb{G}_t$  (at each index $t$), the observation $\mathbf{Y}_t$  follows a multivariate Gaussian distribution whose covariance matrix  diagonalizes in the eigenbasis of the normalized  Laplacian (eqv., the graph-shift operator $\mathbb{A}_{\mathcal{N}}$)  of a (latent) graph. Consequently, the eigenvalues  of the  covariance matrix is of the form $g^{-1}(\boldsymbol{\lambda}_t, \boldsymbol{\theta}_t)$, where $\boldsymbol{\lambda}_t$ denotes the eigenvalues of $\mathbb{L}_{t}$. 
The SGM framework is very versatile and allows for time-varying filter-parameters (i.e., $\boldsymbol{\theta}_t$) and/or time-varying graphs, as well as introducing temporal dependence.

Here we discuss two examples of SGM models which can be viewed as natural extensions of the GAR model proposed in this paper. 
One is to consider a general order graphical autoregressive process:
\begin{equation*}
	\mbf{Y} = (\theta_0 \mathbb{I}_p + \sum_{j=1}^q \theta_j \mathbb{L}^j)^{-1} \mbf{Z}, ~~~ \mbf{Z} \sim N(\mathbf{0}_p, \mathbb{I}_p),
\end{equation*}
where $q \geq 1$ is an integer.  We refer to this as the GAR($q$) model (with the GAR model in this paper being the GAR(1) model).

Another extension is to introduce temporal dependence while requiring the graph-referenced process to be stationary with respect to both time and a graph. Specifically, we have the following \textit{temporal autoregressive- graphical autoregressive (TAR-GAR)} model: 
\begin{eqnarray*}
\mathbf{Y}_t &=& R_1 \mathbf{Y}_{t-1} + \mathbf{U}_t, ~~~ t=2,\cdots, n,\\
\mathbf{U_t} &\sim&_{i.i.d.}  N(\mathbf{0}_p, \Sigma),  ~~~\mathbf{U}_t \independent \mathbf{Y}_{t-1}\\
\mathbf{Y}_1 &\sim& N(\mathbf{0}_p, \Gamma_0).
\end{eqnarray*}
For graph stationarity, we assume the coefficient matrix $R_1$, the covariance $\Sigma$ of the innovation process $\mathbf{U}_t$, and the covariance $\Gamma_0$ of the observation process $\mathbf{Y}_t$ are co-diagonalizable, and 
\begin{eqnarray*}
    R_1&=&\eta_0 \mathbb{I}_p + \eta_1 \mathbb{L}, ~~~ \eta_0>0, ~\eta_1>0, ~~~ \mathbb{L} \in \mathcal{L}_{\mathcal{N},p}\\
    \Sigma&=& (\theta_0\mathbb{I}_p+ \theta_1 \mathbb{L})^{-2}, ~~~ \theta_0>0, ~ \theta_1>0. 
\end{eqnarray*}
In addition, to ensure temporal stationarity, we assume 
$$
|\eta_0+\eta_1 \lambda_j| <1, ~~~ j=1,\cdots,p,
$$
where $\lambda_j$s are eigenvalues of the normalized Laplacian matrix $\mathbb{L}$. 
Note that, in the TAR-GAR model, the innovation process $\mathbf{U}_t$ follows a GAR model.

\appendix
\titleformat{\section}{\normalfont\Large\bfseries}{Appendix \thesection:}{1em}{}
\numberwithin{equation}{section}
\numberwithin{table}{section}
\numberwithin{figure}{section}

\section{Laplacian matrix characterization and GAR model identifiability}
\label{sec: appendix_lemma}

\noindent \textbf{Proof of Lemma \ref{lemma: L_space}}. If $\mbb{L}=\mbb{D}-\mbb{A}$, where  $\mbb{A} \in \mathcal{A}_p$ and  $\mbb{D}$ is a diagonal matrix with the $i$th diagonal entry being $d_i=\sum_{j=1}^p \mbb{A}_{ij}$,  then since $\mbb{A}$ is symmetric,  $\mbb{L}$ is symmetric. Moreover $\mbb{L} \mathbf{1}_p=(\mbb{D}-\mbb{A}) \mathbf{1}_p=\mathbf{0}_p$
 as the $i$th row sum of $\mbb{D}$ and that of $\mbb{A}$ both are equal to $d_i$ for $i=1,\cdots, p$. 
 In addition, for $1 \leq i \not=j  \leq p$, $\mbb{L}_{ij}=-\mbb{A}_{ij} \leq 0$ as off-diagonal entries of $\mbb{D}$ are all zero and off-diagonal entries of $\mbb{A}$ are all non-negative. 
Finally, for any $\mbf{x}=(x_1,\cdots, x_p)^T \in \mbb{R}^p$, it is easy to see that 
$$
\mbf{x}^T \mbb{L} \mbf{x} = \frac{1}{2} \sum_{i=1}^p \sum_{j=1}^p \mbb{A}_{ij} (x_i-x_j)^2 \geq 0 
$$
as all entries of $\mbb{A}$ are non-negative. 
So $\mbb{L}$ is positive semi-definite.

On the other hand, if a $p$ by $p$ matrix $\mbb{L}$ satisfies (i)-(iii), then define a $p$ by $p$ matrix $\mbb{A}$ as 
$$
\mbb{A}_{ij}=\left\{ \begin{array}{ccc}
\text{arbitrary non-negative number}&if& 1 \leq i=j \leq p\\
-\mbb{L}_{ij}&if & 1 \leq i \not= j \leq p
\end{array}
\right.
$$
Then by (i) $\mbb{A}$ is symmetric and by (iii) all entries of $\mbb{A}$ are non-negative, so $\mbb{A} \in \mathcal{A}_p$. 

Now define  $\mbb{D}$ as a diagonal matrix with the $i$th diagonal entry being $d_i=\sum_{j=1}^p \mbb{A}_{ij}$. By (ii) $\mbb{L} \mbf{1}_p =\mbf{0}_p$, we have $\sum_{j=1}^p \mbb{L}_{ij}=0$. Therefore, 
the $i$th diagonal entry of  $\mbb{D}-\mbb{A}$ is equal to $\sum_{j \not=i} \mbb{A}_{ij} =-\sum_{j \not= i} \mbb{L}_{ij}=\mbb{L}_{ii}$ for all $1 \leq i \leq p$. Moreover, the $(i,j)$th off-diagonal entry of the matrix $\mbb{D}-\mbb{A}$ is equal to $-\mbb{A}_{ij}=\mbb{L}_{ij}$ for all $1 \leq i \not= j \leq p$. Therefore,
$\mbb{L}=\mbb{D}-\mbb{A} \in \mathcal{L}_p$. \\

\begin{remark}
    From the above proof, it can be seen that, given a Laplacian matrix $\mbb{L}$, the diagonal entries of the corresponding adjacency matrix $\mbb{A}$ can take arbitrary non-negative values, whereas the off-diagonal entries of $\mbb{A}$ are uniquely determined by $\mbb{L}$.  
\end{remark}

\noindent \textbf{Proof of Lemma \ref{lemma: L_N_space}}. 
If $\mbb{L}= \mbb{I}_p - \mbb{D}^{-1/2}\mbb{A} \mbb{D}^{-1/2}$, where   $\mbb{A} \in \mathcal{A}_p$ with  $d_i=\sum_{j=1}^p \mbb{A}_{ij}>0$  for all $1 \leq i \leq p$ and  $\mbb{D}$ is a diagonal matrix with the $i$th diagonal entry being $d_i$, then since $\mbb{A}$ is symmetric, $\mbb{L}$ is symmetric. 

Now define $\mbf{v}_0=(\sqrt{d_1},\cdots, \sqrt{d_p})^T$. Then $\mbf{v}_0 \succ \mbf{0}$ and 
$$
\mbb{L}\mbf{v}_0=\mbf{v}_0-\mbb{D}^{-1/2}\mbb{A} \mbb{D}^{-1/2}\mbf{v}_0=\mbf{v}_0-\mbb{D}^{-1/2}\mbb{A} \mbf{1}_p=\mbf{v}_0-\mbf{v}_0=\mbf{0}_p.
$$
So (ii) is satisfied. Moreover, for all $1 \leq i \leq p$:
$$
\mbb{L}_{ii} = 1-\frac{\mbb{A}_{ii}}{\sum_{j=1}^p \mbb{A}_{ij}} \in [0,1]
$$
and for all $1 \leq i \not=j \leq p$ 
$$
\mbb{L}_{ij} = -\frac{\mbb{A}_{ij}} {\sqrt{\sum_{l=1}^p \mbb{A}_{il} \sum_{l=1}^p \mbb{A}_{jl}}} \in [-1,0]
$$
as all entries of $\mbb{A}$ are non-negative. So (iii) and (iii)' are satisfied. 

Finally, for any $\mbf{x}=(x_1,\cdots, x_p)^T \in \mbb{R}^p$, it is easy to see that 
$$
\mbf{x}^T \mbb{L} \mbf{x} = \frac{1}{2} \sum_{i=1}^p \sum_{j=1}^p \mbb{A}_{ij} \bigl(\frac{x_i}{\sqrt{d_i}}-\frac{x_j}{\sqrt{d_j}}\bigr)^2 \geq 0 
$$
as all entries of $\mbb{A}$ are non-negative. So $\mbb{L}$ is positive semi-definite.

On the other hand, if a $p$ by $p$ matrix $\mbb{L}$ satisfies (i)-(iii), then define a $p$ by $p$ matrix 
$$
\mbb{A}:=\mbb{D}_v(\mbb{I}_p -\mbb{L})\mbb{D}_v,
$$ 
where 
$\mbb{D}_v$ is a $p$ by $p$ diagonal matrix with the $i$th diagonal entry being $v_i$ -- the $i$th entry of the vector $\mbf{v}_0$ ($1 \leq i \leq p$). By (i), $\mbb{A}$ is symmetric. By (iii) and $\mbf{v}_0 \succ \mbf{0}$, all entries of $\mbb{A}$ are non-negative. So $\mbb{A} \in \mathcal{A}_p$.

By (ii)
$$
\mbb{A}\mbf{1}_p= \mbb{D}_v (\mbb{I}_p -\mbb{L})\mbb{D}_v \mbf{1}_p=\mbb{D}_v (\mbb{I}_p -\mbb{L})\mbf{v}_0=\mbb{D}_v(\mbf{v}_0-\mbb{L}\mbf{v}_0)=\mbb{D}_v \mbf{v}_0=(v_1^2,\cdots, v_p^2)^T. 
$$
So $d_i =\sum_{j=1}^p \mbb{A}_{ij}=v_i^2$ for $1 \leq i \leq p$.  

Now define  $\mbb{D}$ as a diagonal matrix with the $i$th diagonal entry being $d_i$. Then $\mbb{D}_v = \mbb{D}^{1/2}$ and 
$$
\mbb{A}=\mbb{D}^{1/2}(\mbb{I}_p -\mbb{L})\mbb{D}^{1/2}
$$
and thus 
$$
\mbb{L}=\mbb{I}_p - \mbb{D}^{-1/2}\mbb{A} \mbb{D}^{-1/2} \in \mcl{L}_{\mcl{N}, p}.
$$

\begin{remark}
From the above proof, it can be seen that, 
    given a normalized Laplacian matrix $\mbb{L}$, the corresponding adjacency matrix $\mbb{A}$ is not uniquely determined (as $\mbf{v}_0$ is not unique). However, the zero pattern of the off-diagonal entries of $\mbb{A}$ is the same as that of $\mbb{L}$. 
\end{remark}

\noindent \textbf{Proof of Lemma \ref{lemma:identifiability}.}
Suppose two sets of parameters $(\theta_0, \theta_1,\mbb{L})$ and $(\wt{\theta}_0, \wt{\theta}_1, \wt{\mbb{L}})$ in the GAR model (\ref{eq:GAR_1_norm}) parameter space  $\mathbb{R}^{+} \otimes \mathbb{R}^{+} \otimes 	\mcl{L}_{\mcl{N},p}$ have
\begin{equation}
\label{eq: GAR_ident_proof}
\Sigma = (\theta_0 \mathbb{I}_p +\theta_1 \mathbb{L})^{-2} = (\wt{\theta}_0 \mathbb{I}_p +\wt{\theta_1} \wt{\mathbb{L}})^{-2},
\end{equation}
where $\Sigma$ is a positive definite matrix. We  want to show:
$$
\theta_0=\wt{\theta}_0, ~~~ \theta_1 \cdot \mbb{L} = \wt{\theta}_1 \cdot \wt{\mbb{L}}. 
$$

Let $Q$ denote a matrix consisting of a full set of eigenvectors of $\Sigma$. So 
$
Q^T \Sigma Q= \Lambda, 
$
where $\Lambda$ is a diagonal matrix consisting of the corresponding eigenvalues of $\Sigma$. Denote the $i$th diagonal entry of $\Lambda$ by $\lambda_i$ (for $1 \leq i \leq p$).  Without loss of generality, assume $\lambda_1 \geq \lambda_2 \geq \cdots \geq \lambda_p>0$.  

By (\ref{eq: GAR_ident_proof}), $\mbb{L}$ and $\wt{\mbb{L}}$ have the same eigen-subspaces as $\Sigma$, so 
$$
Q^T \mbb{L} Q= D, ~~~ Q^T \wt{\mbb{L}} Q= \wt{D},
$$
where $D$ and $\wt{D}$ are diagonal matrices consisting of eigenvalues of $\mbb{L}$ and $\wt{\mbb{L}}$, respectively. Therefore, by (\ref{eq: GAR_ident_proof})
$$
\Lambda^{-1/2} = \theta_0 \mbb{I}_p + \theta_1 D = \wt{\theta}_0 \mbb{I}_p + \wt{\theta}_1 \wt{D}. 
$$
Denote the $i$th diagonal entry of $D$ by $D_i$.  Then we have 
$$
\lambda_1^{-1/2}=\theta_0+\theta_1 D_1 \leq \lambda_2^{-1/2} = \theta_0+\theta_1 D_2 \leq \cdots \leq \lambda_p^{-1/2}=\theta_0+\theta_1 D_p.
$$
By $\theta_1>0$, we have $D_1 \leq D_2\leq \cdots \leq D_p$. Since $\mbb{L}$ is a p.s.d. matrix with the smallest eigenvalue being zero, $D_1=0$ 
and thus $\theta_0=\lambda_1^{-1/2}$. Similarly, we have $\wt{\theta}_0=\lambda_1^{-1/2}(=\theta_0)$. Therefore,
$
 \theta_1 D = \wt{\theta}_1 \wt{D},
$
which means $\theta_1 Q D Q^T=\wt{\theta}_1 Q \wt{D} Q^T$, i.e., $\theta_1 \cdot \mbb{L}= \wt{\theta}_1 \cdot \wt{\mbb{L}}$.

Moreover, if we assume the graph has no self-loop, then the parameter space for $\mbb{L}$ becomes 
$
\mcl{L}_{\mcl{N},p} \cap \{\mbb{L} \in \mbb{R}^{p \times p}: \mbb{L}_{ii} \equiv 1,~ i=1,\cdots, p\}.
$
Within this space, $\theta_1 \cdot \mbb{L}= \wt{\theta}_1 \cdot \wt{\mbb{L}}$ implies that $\theta_1=\wt{\theta}_1$ and  $\mbb{L}=\wt{\mbb{L}}$.

\section{ADMM algorithms}
\label{sec: appendix_admm}

\subsection{Algorithm for $\mbb{L}$: given  $\theta_0$ and $\mathbf{v}_0$}
\label{subsec:ADMM_sep}

Here we describe the ADMM algorithm for $\mbb{L}$ when $\theta_0$ and $\mathbf{v}_0$ are given and fixed.  The Step 1 estimator (\ref{eq:GAR_step1_estimator}) can be obtained by this algorithm while setting $\theta_0=\wh{\theta}_0^{(ini)}$ and $\mbf{v}_0=\mbf{0}_p$.

By (\ref{eq: GAR_penloglike}), the goal is to minimize the following 
$$
 \frac{1}{2} \tr((\theta_0 \mbb{I}_p + \mbb{L} )^{2}\widehat{\Sigma}) - \log\det(\theta_0 \mbb{I}_p +  \mbb{L}) + \lambda \|\mbb{L}\|_{1,off}
$$
with respect to $\mathbb{L} \in \mbb{R}^{p \times p}$ subject to the  constraints: 
$$
\mbb{L} ~\text{p.s.d.}; ~ \mbb{L} = \mbb{L}^T; ~ \mbb{L}_{ij} \leq 0~\mbox{for all}~ 1 \leq i \neq j \leq p; ~\mbb{L} \mbf{v_0} = \mbf{0}_p
$$

We first notice that the task can be rewritten as:

$$
\min_{\mbb{L},Z, W} \quad \tr\big ((\theta_0\mbb{I}_p + \mbb{L})^2 \hat{\Sigma} \big ) - 2\log\det(\theta_0 \mbb{I}_p + Z) - \lambda \cdot \tr \big( \mbb{L}\tilde{J}_p \big ) + \tilde{I}(W) 
$$
where 
$$
\tilde{I}(W) = 
\begin{cases}
    0, &\text{if} ~ w_{ij} \ge 0 ~\text{for all $i \not=j$  }\\ 
    +\infty, &\text{if otherwise}
\end{cases}
~~~~\text{and}~~ \tilde{J}_p = 2(\mbf{1}_p\mbf{1}_p^T - \mbb{I}_p)
$$ 
subject to:
$$
\quad \mbb{L} - Z =  \mathbf{O}_p, \quad \mbb{L} + W = \mathbf{O}_p, \quad \mbb{L}\mathbf{v}_0 = \mathbf{0}_p, \quad Z \succeq  \mathbf{O}, \quad \mbb{L} = \mbb{L}^{T}. 
$$
In the above, $\mbf{O}_p$ denotes the $p$ by $p$ zero matrix, $\mbf{0}_p$ denotes the p-dimensional zero vector, $\mathbf{1}_p$ denotes  the p-dimensional vector of ones, and $\mbb{I}_p$ denotes the $p$ by $p$ identity matrix.

The scaled augmented Lagrangian is given by:
\begin{eqnarray*}
    L_{\rho}(\mbb{L}, Z, W; U, V, \mbf{r}) &= &
    \tr\big((\theta_0\mbb{I}_p + \mbb{L})^2\wh{\Sigma}\big) - 2\log\det(\theta_0\mbb{I}_p + Z) - \lambda \cdot \tr(\mbb{L}\tilde{J}_p)\\
    &+& \frac{\rho}{2}\|\mbb{L}-Z+U\|^2_2 + \frac{\rho}{2}\|\mbb{L} + W + V\|^2_2 + \frac{\rho}{2} \|\mbb{L}\mbf{v}_0 + \mbf{r}\|^2_2  + \tilde{I}(W), 
\end{eqnarray*}
where $U, V \in \mbb{R}^{p \times p}$ and $\mbf{r} \in \mbb{R}^p$ are \textit{dual variables} and $\rho >0$.  

\begin{remark}
In practice,  we can set $\rho = \max\{\lambda, \epsilon_{\rho}\}$ following the recommendation in \cite{wahlberg2012admm}, where  $\epsilon_{\rho}$ is a small value, e.g., $\epsilon_{\rho}=0.01$.
\end{remark}

Collecting terms in $L_{\rho}(\cdot)$ related to each splitted-variable results in the following functions: 
\begin{eqnarray*}
    F(\mbb{L} \big|Z, W, U, V, \mbf{r}) &=& \tr \big((\theta_0\mbb{I}_p + \mbb{L})^2 \wh{\Sigma}\big) - \lambda \cdot \tr(\mbb{L} \tilde{J}_p) \\
    &+&  \frac{\rho}{2}\|\mbb{L}-Z +U\|^2_2 + \frac{\rho}{2}\|\mbb{L} + W + V\|^2_2 
    + \frac{\rho}{2} \|\mbb{L}\mbf{v}_0 + \mbf{r}\|^2_2\\
    G(Z \big|\mbb{L}, U )&=& -2\log\det(\theta_0\mbb{I}_p+Z) + \frac{\rho}{2}\|\mbb{L}-Z+U\|^2_2 \\ 
    H(W \big|\mbb{L}, V) &=&   \frac{\rho}{2}\|\mbb{L} + W + V\|^2_2 + \tilde{I}(W)
\end{eqnarray*}

The ADMM algorithm  is summarized in algorithm \ref{alg:ADMM_L2}.  

\begin{algorithm}[H]
\caption{ADMM for estimating $\mbb{L}$ given $\theta_0, \mathbf{v}_0$}
\label{alg:ADMM_L2}
    \begin{algorithmic}
        \Input{$\theta_0, \mathbf{v}_0$}
       \Initial{$Z^{(0)} = W^{(0)} = U^{(0)} =V^{(0)}= \mathbf{O}_p, \mbf{r}=\mbf{0}_p; k=0$}        
        \While{$\|\gamma^{(k)}\|_2 > \epsilon^{\text{primal}}$ {\bf or} $\|s^{(k)}\|_2 > \epsilon^{\text{dual}}$} \\
            \CommentLeft{{\it Primal updates}} \\
            \State $\mbb{L}^{(k+1)} = {\arg \min}_{\mbb{L} = \mbb{L}^T}F(\mbb{L} \big|Z^{(k)}, W^{(k)}, U^{(k)}, V^{(k)}, \mbf{r}^{(k)})$
            \State $Z^{(k+1)} = {\arg \min}_{Z \succeq \mathbf{O}} G(Z \big|\mbb{L}^{(k+1)}, U^{(k)})$ 
            \State $W^{(k+1)} = {\arg \min}_{W} H(W \big|\mbb{L}^{(k+1)}, V^{(k)})$ 
            \State \vspace{3mm}
            \CommentLeft{{\it Dual updates}} \\
            \State $U^{(k+1)} = U^{(k)} + (\mbb{L}^{(k+1)} - Z^{(k+1)})$ 
            \State $V^{(k+1)} = V^{(k)} + (\mbb{L}^{(k+1)} + W^{(k+1)})$  
            \State $R^{(k+1)} = R^{(k)} + \mbb{L}^{(k+1)}\mathbf{v}_0$ \\

            \CommentLeft{{\it $k \leftarrow k+1$}} 
        \EndWhile \\
    \Output \text{ $\iter{\mbb{L}}$, $\iter{Z}$, $\iter{W}$}
        
    \end{algorithmic}
\end{algorithm}

In the above, the primal and dual residuals at the $k^{th}$ iteration are defined as:
\begin{eqnarray*}
\|\iter{\gamma}\|^2_2 &=& \|\iter{\mbb{L}} - \iter{Z}\|^2_2 + \|\iter{\mbb{L}} + \iter{W}\|^2_2 + \|\iter{\mbb{L}}\mbf{v}_0\|^2_2  ~~~ (Primal)\\ 
    \|\iter{s}\|^2_2 &=& \rho^2\|\iter{Z} - \itern{Z} + \itern{W} - \iter{W}\|^2_2 ~~~ (Dual).
\end{eqnarray*}
Moreover, let $\epsilon_{abs}>0, \epsilon_{rel} >0$ and define: 
\begin{eqnarray*}
    \epsilon^{primal} &=& \sqrt{p(2p+1)}\epsilon^{abs} + \epsilon^{rel} \max\big\{\sqrt{2\|\iter{\mbb{L}}\|^2_2 + \|\iter{\mbb{L}}\mbf{v}_0\|^2_2}, \sqrt{\|\iter{Z}\|^2_2 + \|\iter{W}\|^2_2} \big\} \\ 
    \epsilon^{dual} &=& p \cdot \epsilon^{abs} + \epsilon^{rel}\cdot\rho\cdot \|\iter{U} + \iter{V} + \iter{\mbf{r}}\mbf{v}_0^T\|_2 
\end{eqnarray*}

The ADMM algorithm stops if and only if $\|\iter{\gamma}\|_2 \le \epsilon^{pri}$ {\em and} $\|\iter{s}\|_2 \le \epsilon^{dual}$.

\begin{remark}
    \text{In practice, we set $\epsilon^{abs} \approx 10^{-5} \sim 10^{-4}$, and $\epsilon^{rel} \approx 10^{-3} \sim 10^{-2}$}.
\end{remark}

The \textit{primal updates} are given as follows. 

    \subsubsection*{L-updates} Note that
    $$
   \iter{F}(\mbb{L}):= F(\mbb{L} \big|Z^{(k)}, W^{(k)}, U^{(k)}, V^{(k)}, \mbf{r}^{(k)}) = \tr\big(C\mbb{L}^2\big) + \tr\big(\iter{D}\mbb{L}\big)
    $$
    where 
    $$
    C = \wh{\Sigma} + \rho \cdot \mbb{I}_p + \frac{\rho}{2}\mbf{v}_0\mbf{v}_0^T
    $$ 
    and 
    $$\iter{D} = 2\theta_0\wh{\Sigma} - \lambda\cdot \tilde{J}_p-\rho(\iter{Z}-\iter{U}) + \rho(\iter{W} + \iter{V}) + \frac{\rho}{2}(\iter{\mbf{r}}\mbf{v}_0^T + \mbf{v}_0(\iter{\mbf{r}})^T).
    $$
    
   The normal equation for $\mbb{L}$-minimization is then:
    $$
    C\mbb{L} + \mbb{L}C + \iter{D} = \mbf{O}_p.
    $$
Since $\iter{D}$ is symmetric, if $\mbb{L}$ satisfies the above equation, then $\mbb{L}^T$ also satisfies the above equation. Since $\iter{F}(\cdot)$ is a convex function in $\mbb{L}$, so the solution of the normal equation is unique  and thus satisfy $\mbb{L}=\mbb{L}^T$.

    Let $Q D Q^T$ denote the eigen-decomposition of $C$. Define $\iter{\tilde{D}} = Q^T\iter{D}Q$,  and $\wt{\mbb{L}} = Q^T\mbb{L}Q$.  Then the normal equation becomes:  
    $$
    D\wt{\mbb{L}} + \wt{\mbb{L}}D + \iter{\tilde{D}} = \mbf{O}_p,
    $$
    where $D$ is a diagonal matrix (consisting of the eigenvalues of $C$). 
    Let $D_x = \mbb{I}_p \otimes D + D \otimes \mbb{I}_p$, where $\otimes$ denotes the Kronecker product.  It is easy to see that 
   $$
    vec(\wt{\mbb{L}}) = -D_x^{-1}vec(\iter{\tilde{D}}),
   $$
where $vec(\cdot)$ is the matrix vectorization operator. Finally, set
$$
\itern{\mbb{L}}: = Q\wt{\mbb{L}}Q^T.
$$

     \subsubsection*{Z-updates} 
     Let $\iter{P} \iter{\Lambda} (\iter{P})^T$ be the eigen-decomposition of $\itern{\mbb{L}} + \iter{U}$. 
     It can be shown that $Z^{(k+1)} = {\arg \min}_{Z \succeq \mathbf{O}} G(Z \big|\mbb{L}^{(k+1)}, U^{(k)})$  takes the form $\iter{P} \itern{\tilde{\Lambda}}(\iter{P})^T$, where $\itern{\tilde{\Lambda}}$ is a diagonal matrix with non-negative diagonals  $\itern{\tilde{\lambda}}_1, \cdots, \itern{\tilde{\lambda}}_p$ that minimize the following function:
    $$
    G(\tilde{\Lambda}) := -2\sum_{j=1}^p\log(\theta_0 + \tilde{\lambda}_j) + \frac{\rho}{2}\sum_{j=1}^p(\tilde{\lambda}_j - \iter{\lambda}_j)^2, ~~ \tilde{\lambda}_j \geq 0, ~ j=1,\cdots, p, 
    $$  
    where $\iter{\lambda}_j$ is the $j$th diagonal entry of the  diagonal matrix $\iter{\Lambda}$ (consisting of eigenvalues of $\itern{\mbb{L}} + \iter{U}$). 
    
    This leads to:
    $$
    \itern{\tilde{\lambda}}_j = \max \big\{0, \lambda^{*,+}_j - \theta_0\big\}, ~~~j=1,\cdots, p,
    $$
where 
$$
\lambda_j^{*,+} = \frac{\rho(\theta_0 + \iter{\lambda}_j) + \sqrt{\rho^2(\theta_0+\iter{\lambda}_j)^2 + 8\rho}}{2\rho}.
$$

\subsubsection*{W-updates}  It is easy to see that the minimizer for $ H(W \big|\mbb{L}^{(k+1)}, V^{(k)})$ has the $(i,j)$th entry:  
    \begin{align*}        
    \itern{W}_{ij} = \begin{cases}
        -(\itern{\mbb{L}}_{ii} + \iter{V}_{ii}), & i = j \\ 
        \max\big\{0, -(\itern{\mbb{L}}_{ij} + \iter{V}_{ij}) \big\}, & i \ne j
    \end{cases}
    \end{align*}

\begin{remark}
\label{remark:alg_null_set}
Given the null-set $\mathcal{N}=\mathcal{N}_{\lambda, \epsilon_{thre}}$ (\ref{eq:null_set}), to obtain the Step 2 estimator (\ref{eq:GAR_step2_estimator}), we only need to set $\lambda=0$ in  Algorithm \ref{alg:ADMM_L2} and modify the W-updates as follows:

\noindent \textbf{W-updates given the null-set $\mathcal{N}$:}  \\
    \begin{align*}        
    \itern{W}_{ij} = \begin{cases}
        -(\itern{\mbb{L}}_{ii} + \iter{V}_{ii}), & i = j \\ 
        \max\big\{0, -(\itern{\mbb{L}}_{ij} + \iter{V}_{ij}) \big\}, & i \ne j ~\&~ (i,j) \notin  \mathcal{N}\\
        0, & i \ne j ~\&~ (i,j) \in  \mathcal{N}
    \end{cases}
    \end{align*}
    
\end{remark}

\subsection{Algorithm for $(\theta_0, \mbb{L})$: given   $\mbf{v}_0$}
\label{subsec:ADMM_simu}

Next, we describe the ADMM algorithm for $(\theta_0, \mbb{L})$ when $\mbf{v}_0$ is given and fixed.  

Our goal is to minimize the following 
$$
 \frac{1}{2} \tr((\theta_0 \mbb{I}_p + \mbb{L} )^{2}\widehat{\Sigma}) - \log\det(\theta_0 \mbb{I}_p +  \mbb{L}) + \lambda \|\mbb{L}\|_{1,off}
$$
with respect to $(\theta_0, \mathbb{L}) \in \mathbb{R}^{+} \times \mbb{R}^{p \times p}$ subject to the  constraints: 
$$
\mbb{L} ~\text{p.s.d.}; ~ 
\mbb{L} = \mbb{L}^T; ~ \mbb{L}_{ij} \leq 0~\mbox{for all}~ 1 \leq i \neq j \leq p; ~\mbb{L} \mbf{v_0} = \mbf{0}_p
$$

This task can be re-written as:

\[
\min_{\mbb{L},Z, W, \theta_0, \phi} \quad \tr\big ((\theta_0\mbb{I}_p + \mbb{L})^2 \hat{\Sigma} \big ) - 2\log\det(\phi \mbb{I}_p + Z) - \lambda \cdot \tr \big( \mbb{L}\tilde{J}_p \big ) + \tilde{I}(W) 
\]

where 
$$
\tilde{I}(W) = 
\begin{cases}
    0, &\text{if} ~ w_{ij} \ge 0 ~\text{for all $i \not=j$  }\\ 
    +\infty, &\text{if otherwise}
\end{cases}
~~~~\text{and}~~ \tilde{J}_p = 2(\mbf{1}_p\mbf{1}_p^T - \mbb{I}_p)
$$ 
subject to:
$$
\quad \mbb{L} - Z =  \mathbf{O}_p, \quad \mbb{L} + W = \mathbf{O}_p, \quad \mbb{L}\mathbf{v}_0 = \mathbf{0}_p, \quad Z \succeq  \mathbf{O}, \quad \mbb{L} = \mbb{L}^{T}
$$

and 

$$
\theta_0 - \phi = 0, \quad \phi > 0.
$$

This results in a scaled augmented Lagrangian:

\begin{eqnarray*}
    L_{\rho}(\mbb{L}, Z, W, \theta_0, \phi; U, V, \mbf{r}, t) &= &
    \tr\big((\theta_0\mbb{I}_p + \mbb{L})^2\wh{\Sigma}\big) - 2\log\det(\phi\mbb{I}_p + Z) - \lambda \cdot \tr(\mbb{L}\tilde{J}_p) \\
    &+& \frac{\rho}{2}\|\mbb{L}-Z+U\|^2_2 + \frac{\rho}{2}\|\mbb{L} + W + V\|^2_2 + \frac{\rho}{2} \|\mbb{L}\mbf{v}_0 + \mbf{r}\|^2_2  \\ 
    &+& \frac{\rho}{2}(\theta_0-\phi+t)^2 + \tilde{I}(W), 
\end{eqnarray*}
where $U, V \in \mbb{R}^{p \times p}$, $\mbf{r} \in \mbb{R}^p$ and $t \in \mbb{R}$ are \textit{dual variables} and $\rho >0$.

Collecting terms in $L_{\rho}(\cdot)$ related to each splitted-variable results in the following functions: 
\begin{eqnarray*}
    F(\mbb{L}, \theta_0 \big|Z, W, \phi, U, V, \mbf{r}, t) &=& \tr \big((\theta_0\mbb{I}_p + \mbb{L})^2 \wh{\Sigma}\big) - \lambda \cdot \tr(\mbb{L} \tilde{J}_p) \\
    &+&  \frac{\rho}{2}\|\mbb{L}-Z +U\|^2_2 + \frac{\rho}{2}\|\mbb{L} + W + V\|^2_2 
    + \frac{\rho}{2} \|\mbb{L}\mbf{v}_0 + \mbf{r}\|^2_2\\
     &+& \frac{\rho}{2}(\theta_0-\phi+t)^2 \\
    G(Z, \phi \big|\mbb{L}, \theta_0, U, t)&=& -2\log\det(\phi\mbb{I}_p+Z) + \frac{\rho}{2}\|\mbb{L}-Z+U\|^2_2 +\frac{\rho}{2}(\theta_0-\phi+t)^2 \\ 
    H(W \big|\mbb{L}, V) &=&  \frac{\rho}{2}\|\mbb{L} + W + V\|^2_2 +\tilde{I}(W)
\end{eqnarray*}

The ADMM algorithm is summarized in Algorithm \ref{alg:ADMM_Lap}.

\begin{algorithm}[H]
\caption{ADMM for estimating $(\theta_0, \mbb{L})$ given $\mathbf{v}_0$}
\label{alg:ADMM_Lap}
    \begin{algorithmic}
        \Input{$\mathbf{v}_0$}
       \Initial{$Z^{(0)} = W^{(0)} = U^{(0)} =V^{(0)}= \mathbf{O}_p, \mbf{r}=\mbf{0}_p; \phi = \epsilon; k=t=0$}        
        \While{$\|\gamma^{(k)}\|_2 > \epsilon^{\text{primal}}$ {\bf or} $\|s^{(k)}\|_2 > \epsilon^{\text{dual}}$} \\
            \CommentLeft{{\it Primal updates}} \\
            \State $(\theta_0^{(k+1)}, \mbb{L}^{(k+1)}) = {\arg \min}_{\theta_0, \mbb{L} = \mbb{L}^T}F(\mbb{L}, \theta_0 \big|Z^{(k)}, W^{(k)}, \phi^{(k)}, U^{(k)}, V^{(k)}, \mbf{r}^{(k)}, t^{(k)})$
            \State $(\phi^{(k+1)}, Z^{(k+1)}) = {\arg \min}_{\phi \geq \epsilon, Z \succeq \mathbf{O}} G(Z, \phi \big|\mbb{L}^{(k+1)}, \theta_0^{(k+1)}, U^{(k)}, t^{(k)})$ 
            \State $W^{(k+1)} = {\arg \min}_{W} H(W \big|\mbb{L}^{(k+1)}, V^{(k)})$ 
            \State \vspace{3mm}
            \CommentLeft{{\it Dual updates}} \\
            \State $U^{(k+1)} = U^{(k)} + (\mbb{L}^{(k+1)} - Z^{(k+1)})$ 
            \State $V^{(k+1)} = V^{(k)} + (\mbb{L}^{(k+1)} + W^{(k+1)})$  
            \State $R^{(k+1)} = R^{(k)} + \mbb{L}^{(k+1)}\mathbf{v}_0$ 
            \State $t^{(k+1)} = t^{(k)} + (\theta_0^{(k+1)} - \phi^{(k+1)})$ \\

            \CommentLeft{{\it $k \leftarrow k+1$}} 
        \EndWhile \\
    \Output \text{ $\iter{\mbb{L}}$, $\iter{Z}$, $\iter{W}, \iter{\theta}_0, \iter{\phi}$}
        
    \end{algorithmic}
\end{algorithm}
\begin{remark} In Algorithm \ref{alg:ADMM_Lap}, $\epsilon>0$ is a small value, e.g., $\epsilon=10^{-6}$, to ensure numerically $\phi\geq \epsilon >0$. 
\end{remark}

In the above, the primal and dual residuals at the $k^{th}$ iteration are defined as:
\begin{eqnarray*}
\|\iter{\gamma}\|^2_2 &=& \|\iter{\mbb{L}} - \iter{Z}\|^2_2 + \|\iter{\mbb{L}} + \iter{W}\|^2_2 + \|\iter{\mbb{L}}\mbf{v}_0\|^2_2 + (\iter{\theta}_0 - \iter{\phi})^2  ~~~ (Primal)\\ 
    \|\iter{s}\|^2_2 &=& \rho^2\|\iter{Z} - \itern{Z} + \itern{W} - \iter{W}\|^2_2 + \rho^2(\iter{\phi} - \itern{\phi})^2 ~~~ (Dual).
\end{eqnarray*}
Moreover, let $\epsilon_{abs}>0, \epsilon_{rel} >0$ and define: 
\begin{eqnarray*}
    \epsilon^{primal} &=& 
    \sqrt{p(2p+1)+1}\epsilon^{abs} \\ 
    &+& \epsilon^{rel} \max\bigl\{\sqrt{2\|\iter{\mbb{L}}\|^2_2 + \|\iter{\mbb{L}}\mbf{v}_0\|^2_2 + (\iter{\theta}_0)^2}, \sqrt{\|\iter{Z}\|^2_2 + \|\iter{W}\|^2_2 + (\iter{\phi})^2} \bigr\} \\ 
    \epsilon^{dual} &=& \sqrt{p^2 + 1} \cdot \epsilon^{abs} + \epsilon^{rel}\cdot\rho\cdot\sqrt{\|\iter{U} + \iter{V} + \iter{\mbf{r}}\mbf{v}_0^T\|^2_2 + (\iter{t})^2}
\end{eqnarray*}
The ADMM algorithm stops if and only if $\|\iter{\gamma}\|_2 \le \epsilon^{pri}$ {\em and} $\|\iter{s}\|_2 \le \epsilon^{dual}$.

The primal updates are given as follows.
\subsubsection*{$(\theta_0, \mbb{L})$-updates}
Let $F^{(k)}(\mathbb{L},\theta_0):=F(\mbb{L}, \theta_0 \big|Z^{(k)}, W^{(k)}, \phi^{(k)}, U^{(k)}, V^{(k)}, \mbf{r}^{(k)}, t^{(k)})$. Note that 
$$
F^{(k)}(\mathbb{L},\theta_0) = \tr \big( C \mbb{L}^2\big) + \tr \big(\iter{D}(\theta_0)\mbb{L} \big) + \theta_0^2\cdot (\tr(\hat{\Sigma}) + \frac{\rho}{2}) - \rho\cdot \theta_0\cdot (\iter{\phi} - \iter{t}),
$$
where 
$$
C = \hat{\Sigma} + \rho \mbb{I}_p + \frac{\rho}{2}\mbf{v}_0\mbf{v}_0^T
$$
and
$$
\iter{D}(\theta_0) = 2\theta_0\hat{\Sigma} - \lambda\tilde{J}_p - \rho(\iter{Z} - \iter{U}) + \rho(\iter{W} + \iter{V}) + \frac{\rho}{2} \big(\iter{\mbf{r}}\mbf{v}_0^T + \mbf{v}_0(\iter{\mbf{r}})^T \big)
$$

The normal equations for $(\theta_0,  \mbb{L})$-minimization are then:
    \begin{eqnarray*}
          C\mbb{L} + \mbb{L}C + \iter{D}(\theta_0) &=& \mbf{O}_p \\
          (2\tr(\hat{\Sigma}) + \rho)\cdot \theta_0 + 2\tr(\hat{\Sigma}\mbb{L}) - \rho\cdot (\iter{\phi} - \iter{t}) &=& 0
    \end{eqnarray*}

    Let $Q D Q^T$ denote the eigen-decomposition of $C$. Define  $\wt{\mbb{L}} = Q^T\mbb{L}Q$, and $\iter{\tilde{D}}(\theta_0) = Q^T\iter{D}(\theta_0)Q = 2\theta_0\wt{\Sigma} + \iter{E}$,  where
    $
    \wt{\Sigma} = Q^T\hat{\Sigma}Q
    $
    and
    $$
    \iter{E} = Q^T\bigl\{ -\lambda\tilde{J}_p -\rho(\iter{Z} - \iter{U}) + \rho(\iter{W} + \iter{V}) + \frac{\rho}{2}(\iter{\mbf{r}}\mbf{v}_0^T + \mbf{v}_0(\iter{\mbf{r}})^T) \bigr\}Q.
    $$
    Then the normal equations become:  
    \begin{eqnarray*}
    D\wt{\mbb{L}} + \wt{\mbb{L}}D + 2\theta_0\wt{\Sigma} + \iter{E} &=& \mbf{O}_p \\ 
    (2\tr(\wt{\Sigma}) + \rho)\cdot \theta_0 + 2\tr(\wt{\Sigma}\wt{\mbb{L}}) - \rho\cdot (\iter{\phi} - \iter{t}) &=& 0,
    \end{eqnarray*}
    where $D$ is a diagonal matrix (consisting of the eigenvalues of $C$). 
    Let $D_x = \mbb{I}_p \otimes D + D \otimes \mbb{I}_p$, $S_x = 2vec(\wt{\Sigma})$, and $\delta_x = 2\tr(\hat{\Sigma}) + \rho$, where $\otimes$ denotes the Kronecker product and $vec(\cdot)$ denotes the matrix vectorization operator.  It is easy to see that 

    $$
    \begin{pmatrix}
        vec(\wt{\mbb{L}}) \\ 
        \itern{\theta}_0
    \end{pmatrix}
    =
    \begin{pmatrix}
        D_x & S_x \\ 
        S_x^T & \delta_x
    \end{pmatrix}^{-1}
    \begin{pmatrix}
        -vec(\iter{E}) \\ 
        \rho\cdot (\iter{\phi} - \iter{t})
    \end{pmatrix}.
    $$
Finally, set
$$
\itern{\mbb{L}}: = Q\wt{\mbb{L}}Q^T.
$$

\subsubsection*{$(\phi, Z)$-updates}

 Let $\iter{P} \iter{\Lambda} (\iter{P})^T$ be the eigen-decomposition of $\itern{\mbb{L}} + \iter{U}$. 
     It can be shown that given $\phi \geq \epsilon>0$, $Z^{(k+1)} = {\arg \min}_{Z \succeq \mathbf{O}} G(Z, \phi \big| \mbb{L}^{(k+1)}, \theta_0^{(k+1)}, U^{(k)}, t^{(k)})$  takes the form $\iter{P} \itern{\tilde{\Lambda}}(\iter{P})^T$, where $\itern{\tilde{\Lambda}}$ is a diagonal matrix with non-negative diagonals  $\itern{\tilde{\lambda}}_1, \cdots, \itern{\tilde{\lambda}}_p$.  The $(\phi, Z)$ objective function can then be re-parameterized as follows:
    \begin{eqnarray*}
    G(\tilde{\Lambda}, \phi) := -2\sum_{j=1}^p\log(\phi + \tilde{\lambda}_j) + \frac{\rho}{2}\sum_{j=1}^p(\tilde{\lambda}_j - \iter{\lambda}_j)^2 + \frac{\rho}{2}(\itern{\theta_0} - \phi - \iter{t})^2, ~~ \\ 
    \phi \geq \epsilon,~~~\tilde{\lambda}_j \geq 0, ~ j=1,\cdots, p, 
    \end{eqnarray*}
    where $\iter{\lambda}_j$ is the $j$th diagonal entry of the  diagonal matrix $\iter{\Lambda}$ (consisting of eigenvalues of $\itern{\mbb{L}} + \iter{U}$). 
    
    Equivalently, we can write the objective function as
    $$
    G^*(\Lambda^*, \phi) := -2\sum_{j=1}^p\log(\lambda_j^*) + \frac{\rho}{2}\sum_{j=1}^p(\lambda^*_j - \phi - \iter{\lambda}_j)^2 + \frac{\rho}{2}(\phi - (\itern{\theta}_0 + \iter{t}))^2,
    $$
where 
$
\lambda_j^{*} = \phi + \tilde{\lambda}_j \geq \phi$ for $j=1,\cdots, p$ and $ \phi \geq \epsilon>0
$.

We solve this problem using an alternating-update approach.  This inner-algorithm is summarized in Algorithm \ref{alg:Z_phi_update}.  

\begin{algorithm}[H]
\caption{$(\phi, Z)$-updates}
\label{alg:Z_phi_update}
    \begin{algorithmic}
        \Input{$\iter{\phi},  \iter{\wt{\Lambda}};  \iter{P}, \iter{\Lambda}; \itern{\theta}_0, \iter{t}$ and $\rho>0, \epsilon>0$}
       \Initial{$\phi^{(c)} = \iter{\phi}, \lambda^{*, (c)}_{j} = \phi^{(c)} + \tilde{\lambda}_{j}^{(k)}$}        
        \While{\bf $r^{(new)} > \epsilon_{(new)}$} \\
            \CommentLeft{{\it Alternating updates}} \\
            \State $\phi^* = \frac{1}{p+1}\big( \sum_{j=1}^p (\lambda_{j}^{*, (c)} - \iter{\lambda}_j) + \itern{\theta}_0 + \iter{t} \big)$ 
            \State $\phi^{(new)} = \max \big\{\epsilon, \phi^* \big\}$ \\
            For $j=1,\cdots,p$:\\
            \State $\lambda_{j}^{*,+} = \frac{\rho(\phi^{(new)} + \iter{\lambda}_j) + \sqrt{\rho^2(\phi^{(new)} + \iter{\lambda}_j)^2 + 8\rho}}{2\rho}$
            \State $\lambda_{j}^{*,(new)} = \max \big\{ \phi^{(new)}, \lambda^{*, +}_j\big\}$ \\

            \CommentLeft{{\it $\phi^{(c)} \leftarrow \phi^{(new)}$}; $\lambda_j^{*,(c)} \leftarrow \lambda_j^{*,(new)}, ~j=1,\cdots, p$} 
         
        \EndWhile \\

    \CommentLeft{{\it Re-parameterize}} \\
    \State    $\itern{\tilde{\lambda}}_j = \lambda^{*, (c)}_{j} - \phi^{(c)}$
    \State $\itern{Z} = \iter{P} \wt{\Lambda}^{(k+1)} (\iter{P})^{T}$
    \State $\itern{\phi} = \phi^{(c)}$ \\
    \Output \text{ $\itern{\phi}$, $\itern{Z}$}

    \end{algorithmic}
\end{algorithm}
\textbf{Stopping Criterion.}
We adopt an ADMM-type stopping criterion.  Let $\delta_{abs}>0, \delta_{rel} > 0$ be two small values.   
For each iteration, we define the residual as

\[
r^{(new)} = \sqrt{(\phi^{(new)} - \phi^{(c)})^2 + \|\Lambda^{*,(new)} - \Lambda^{*,(c)}\|_F^2}
\]
where $\Lambda^{*,(c)}$ is a diagonal matrix consisting of $\lambda^{*, (c)}_j, \quad j = 1, \dots, p$, and 

\[
\epsilon_{(new)} = \delta_{abs}\sqrt{p+1} + \delta_{rel}\max\big\{ \sqrt{(\phi^{(new)})^2 + \|\Lambda^{*,(new)}\|_F^2},  \sqrt{(\phi^{(c)})^2 + \|\Lambda^{*,(c)}\|_F^2} \big\}.
\]
Algorithm \ref{alg:Z_phi_update} stops if and only if $r^{(new)} \le \epsilon_{(new)}$.  

\begin{remark}
    \text{In practice, we set $\delta_{abs} \approx 10^{-5} \sim 10^{-4}$ and $\delta_{rel} \approx 10^{-3} \sim 10^{-2}$}.
\end{remark}

\subsubsection*{W-updates} The minimizer for $ H(W \big|\mbb{L}^{(k+1)}, V^{(k)})$ has the $(i,j)$th entry:  
    \begin{align*}        
    \itern{W}_{ij} = \begin{cases}
        -(\itern{\mbb{L}}_{ii} + \iter{V}_{ii}), & i = j \\ 
        \max\big\{0, -(\itern{\mbb{L}}_{ij} + \iter{V}_{ij}) \big\}, & i \ne j
    \end{cases}
    \end{align*}

\begin{remark}
The Step 3 estimator (\ref{eq: GAR_step3_estimator_v0_known}) can be obtained by Algorithm \ref{alg:ADMM_Lap}
while setting $\lambda=0$, $\mbf{v}_0$ to $\widehat{\mbf{v}}_{\lambda, \epsilon_{thre}}$, and modifying the W-updates as in Remark \ref{remark:alg_null_set} given the null-set  $\mathcal{N}=\mathcal{N}_{\lambda, \epsilon_{thre}}$ from Step 1. 

\end{remark}

\subsection{Algorithm  for $\mbf{v}_0$: given $\mbb{L}$}
\label{subsec:ADMM_v0}

Here, we describe a more general algorithm for learning an eigenvector with all positive entries given a positive semidefinite matrix $S \in \mbb{R}^{p \times p}$ and an eigenvalue $\lambda^*$. Given  $\mbb{L}$, an estimator for  $\mbf{v}_0$ can then be obtained by setting $\lambda^* = 0$ and $S = \mbb{L}$.   

Let $\epsilon$ be some pre-specified small value, e.g., $\epsilon=10^{-6}$.  The goal is to minimize the following

\[
\frac{1}{2} \|S\mbf{v} - \lambda^*\mbf{v}\|^2_2
\]

with respect to $\mbf{v} \in \mbb{R}^{p}$ subject to the constraints:

\[
\mbf{v}^T\mbf{v} = 1, \quad \mbf{v}_{i} \ge \epsilon, \quad i = 1, \dots, p.
\]

This task can be re-written as:
\[
\min_{\mbf{v},\mbf{w}} \frac{1}{2} \| S\mbf{v} - \lambda^*\mbf{v}\|^2_2 + \tilde{I}(\mbf{w}),
\]
where 

$$
\tilde{I}(w) = 
\begin{cases} 
0, & \mbf{w}_i \ge \epsilon \text{ for all } i = 1, \dots, p \\ 
+\infty, & \text{if otherwise}
\end{cases}
$$
subject to:
$$
\mbf{v} - \mbf{w} = \mbf{0} \quad \text{and} \quad \mbf{w}^T\mbf{w} = 1.
$$

This results in a scaled augmented Lagrangian given by:
\begin{eqnarray*}
    L_{\rho}(\mbf{v}, \mbf{w}, \mbf{u}) = \frac{1}{2} \|S\mbf{v} - \lambda^*\mbf{v} \|^2_2  + \frac{\rho}{2} \|\mbf{v} - \mbf{w} - \mbf{u} \|^2_2 + \tilde{I}(\mbf{w}),
\end{eqnarray*}
where $\mbf{u} \in \mbb{R}^p$ is a dual variable, and $\rho>0$.  

Let 
\begin{eqnarray*}
F(\mbf{v} \big| \mbf{w, u}) := \frac{1}{2}\mbf{v}^TC\mbf{v} + \frac{\rho}{2}\mbf{v}^T\mbf{v} - \mbf{v}^T\mbf{t} \\ 
H(\mbf{w} \big | \mbf{v}, u) := \tilde{I}(\mbf{w}) + \frac{\rho}{2}\|\mbf{v} - \mbf{w} + \mbf{u} \|^2_2
\end{eqnarray*}
where $C = (S - \lambda^{\ast} \cdot \mbb{I}_p)^2$ and $\mbf{t} = \rho(\mbf{w} - \mbf{u})$.

The algorithm is summarized in Algorithm \ref{alg:ADMM_deg} below.  
\begin{algorithm}[H]
\caption{ADMM for estimating $\mbf{v}$ given $S, \lambda^*$}
\label{alg:ADMM_deg}
    \begin{algorithmic}
        \Input{$S, \lambda^*, \epsilon$}
       \Initial{$\mbf{w}^{(0)}_1 = \dots = \mbf{w}^{(0)}_p = \epsilon$, $\mbf{u}^{(0)} = \mbf{0}$}        
        \While{\bf $\|\gamma^{(k)}\|_2 > \epsilon^{\text{primal}}$ {\bf or} $\|s^{(k)}\|_2 > \epsilon^{\text{dual}}$} \\
            \CommentLeft{{Primal updates}} \\
            \State $\itern{\mbf{v}} = \arg\min_{\mbf{v}^T\mbf{v} = 1}F(\mbf{v} \big | \iter{\mbf{w}}, \iter{\mbf{u}})$
            \State $\itern{\mbf{w}} = \arg\min_{\mbf{w}} H(\mbf{w} \big | \itern{\mbf{v}}, \iter{\mbf{u}})$ \\ 

            \CommentLeft{Dual update}  \\
            \State $\itern{\mbf{u}} = \iter{\mbf{u}} + (\itern{\mbf{v}} - \itern{\mbf{w}})$\\ 
            
            \CommentLeft{$k \leftarrow k + 1$}
        \EndWhile \\

    \Output $\iter{\mbf{v}}, \iter{\mbf{w}}$
    \end{algorithmic}
\end{algorithm}

The primal and dual residuals at the $k^{th}$ iteration are defined as:

\begin{eqnarray*}
\|\iter{\gamma}\|_2 = \|\iter{\mbf{v}} - \iter{\mbf{w}}\|_2 & (primal) \\ 
\|\iter{s}\|_2 = \rho \|\iter{\mbf{w}} - \itern{\mbf{w}} \|_2 & (dual)
\end{eqnarray*}

Let $\epsilon_{abs} > 0$ and $\epsilon_{rel} > 0$ and define:

\begin{eqnarray*}
    \epsilon^{primal} &=& \sqrt{2p} \cdot \epsilon^{abs} + \epsilon^{rel} \max \big\{\|\iter{\mbf{v}}\|_2, \|\iter{\mbf{w}}\|_2 \big\} \\ 
    \epsilon^{dual} &=& \sqrt{p} \cdot \epsilon^{abs} + \epsilon^{rel} \cdot \rho \cdot \|\iter{\mbf{u}}\|_2
\end{eqnarray*}

The ADMM algorithm stops if and only if $\|\iter{\gamma}\|_2 \le \epsilon^{pri}$ {\em and} $\|\iter{s}\|_2 \le \epsilon^{dual}$.

The primal updates are given as follows. 

\subsubsection*{$\mbf{v}$-updates}

Under the constraint $\mbf{v}^T\mbf{v} = 1$, we have:

\[
F(\mbf{v} \big| \mbf{\iter{w}, \iter{u}}) := \frac{1}{2}\mbf{v}^TC\mbf{v} + \frac{\rho}{2} - \mbf{v}^T\mbf{\iter{t}}.
\]

The resulting Lagrangian is:

\[
F(\mbf{v},\mu \big| \mbf{\iter{t}}) = \frac{1}{2} \mbf{v}^TC\mbf{v} - \mbf{v}^T\iter{\mbf{t}} + \mu (\mbf{v}^T\mbf{v} - 1),
\]
where $\mu>0$ is the Lagrange multiplier. 

This yields the following gradient equations:

\begin{eqnarray*}
    \frac{\partial F}{\partial \mbf{v}} &=&(C + 2\mu\mbf{I}_p)\mbf{v} - \iter{\mbf{t}} = \mathbf{0} \\ 
    \frac{\partial F}{\partial \mu} &=& \mbf{v}^T\mbf{v} - 1 = 0
\end{eqnarray*}

Define
$$
\wt{C}(\mu) = C + 2\mu\mbb{I}_p = (S - \lambda^*\mbb{I}_p)^2 + 2\mu\mbb{I}_p.
$$
Let $Q^T \Lambda Q$ denote the eigendecomposition of $S$. So
$$
\wt{C}(\mu) = Q^T \wt{\Lambda}(\mu)Q,
$$
where 
$$\wt{\Lambda}(\mu) = (\Lambda - \lambda^*\mbb{I}_p)^2+2\mu\mbb{I}_p$$

This implies that

$$
Q\mbf{v}(\mu) = \wt{\Lambda}^{-1}(\mu)Q\iter{\mbf{t}}.
$$
Now, let 
$$
\wt{\mbf{v}}(\mu) = Q\mbf{v}(\mu) \quad \text{and} \quad \tilde{\mbf{t}} = Q\iter{\mbf{t}}.
$$
Then since $\|\tilde{\mbf{v}}(\mu)\|_2^2 =\|\mbf{v}(\mu)\|_2^2 = 1$, it can then be shown that $\mu$ satisfies:
$$
\sum_{i=1}^p\frac{(\tilde{t_i})^2}{\big((\lambda_i - \lambda^*)^2 + 2\mu \big)^2} = 1,
$$
which can be solved numerically using a root-solving method.

Once the value of $\mu$ is found, the update for $\mbf{v}$ is given by:
\[
\itern{\mbf{v}} = Q^T \wt{\Lambda}^{-1}(\mu)\tilde{\mbf{t}}
\]

\subsubsection*{$\mbf{w}$-updates}

It is easy to see that the minimizer for $H(\mbf{w}\big| \itern{\mbf{v}}, \iter{\mbf{u}})$ satisfies for $i=1,\cdots p$

$$
w_i = 
\begin{cases}
    \itern{v}_i + \iter{u}_i, & \text{if $\itern{v}_i + \iter{u}_i \ge \epsilon$} \\ 
    \epsilon, & \text{if otherwise}
\end{cases}
$$

\section{Proof details of the theorems}
\label{sec: appendix_proof}

\subsection{Proof of Theorem \ref{thm:consistency_hat_L_theta_0_known}}

We assume throughout that $\mbm{L}  \in \wt{\mcl{L}}_{\mcl{N},p}^+$ (\ref{eq: GAR_step1_space}), the relaxed parameter space by dropping the $\mbf{v}_0$ requirement.

Define
\begin{equation}\label{eq:perturbation_space}
	\mcl{D}_p = \{\Delta \in \mbm{R}^{p\times p}: \Delta = \Delta^T\}. 
\end{equation}

Let 
\begin{eqnarray}
G_\lambda(\Delta) &:=& g_\lambda(\theta_0,\mbm{L}^* + \Delta) - g_\lambda(\theta_0,\mbm{L}^*) \nonumber\\
&=& \ell_0(\mbm{L}^* +\Delta) - \ell_0(\mbm{L}^*) + \lambda( \|\mbm{L}^* + \Delta\|_{1,off} - \|\mbm{L}\|_{1,off}),
\end{eqnarray}
where $\ell_0(\mbm{L}^*) = \ell(\theta_0,\mbm{L}^*)$. Also, let
\[
\mcl{D}_p(\mbm{L}^*) := \{\Delta \in \mcl{D}_p : \mbm{L}^* + \Delta \in \wt{\mcl{L}}_{\mcl{N},p}^+\}.
\]
Note that, this requires in particular that $\Delta_{ij} \leq 0$ whenever $L_{ij}^*=0$.

Then, the estimate $\wh{\mbm{L}}_\lambda = \mbm{L}^* + \wh{\Delta}_\lambda$ satisfies 
\[
\wh{\Delta}_\lambda = \arg\min_{\Delta \in \mcl{D}_p(\mbm{L}^*)} G_\lambda(\Delta).
\]

Define, for any $r > 0$, 
\[
\mcl{B}_p^b(r) := \{\Delta \in \mbm{R}^{p\times p} : \|\Delta \|_F = r\}.
\]
We want to show that, with
\[
r_n := M \sqrt{\frac{(s + p) \log p}{n}},
\]
for $M>0$ being sufficiently large, 
we have with probability tending to 1 (henceforth, \textit{w.p. $\to 1$}),
\begin{equation}
\inf_{\Delta \in \mcl{D}_p(\mbm{L}^*)\cap \mcl{B}_p^b(r_n)} G_{\lambda_n}(\Delta) > G_{\lambda_n}(O_p) = 0,
\end{equation}
where $O_p$ is the $p\times p$ matrix of zeros. This will imply that, \textit{w.p. $\to 1$},  there is a 
local minimum of $G_{\lambda_n}(\Delta)$ within a Frobenius norm neighborhood of radius $r_n$. 

Since the loss function is strictly convex in $\mbm{L}$, the estimator  $\wh{\mbm{L}}_{\lambda}^0$ 
is the unique  minimum of $g_{\lambda_n}(\theta_0,\mbm{L})$ over $\mbm{L} \in \mcl{L}_p$. This proves Theorem \ref{thm:consistency_hat_L_theta_0_known}.

Let $\mbm{L}_{\mcl{S}}^{-}$ denotes the matrix whose $(i,j)$-th entry is $L_{ij} \mbf{1}(i \neq j ~\mbox{and}~ L_{ij}<0)$. Also, let $\bar{\mcl{S}} = \{(i,j): 1\leq i \neq j \leq p ~\mbox{such that}~ L_{ij} = 0\}$.
Define $\Delta_{\mcl{S}}^{-}$ and  $\Delta_{\bar{\mcl{S}}}^{-}$ to be the matrices with $\Delta = \Delta^T$, whose $(i,j)$-th entry is $\Delta_{ij} \mbf{1}(i \neq j ~\mbox{and}~ L_{ij}<0)$, and $\Delta_{ij} \mbf{1}(i \neq j ~\mbox{and}~ L_{ij}=0)$, respectively. Finally, $\Delta^{+}$ is the diagonal $p\times p$ matrix whose diagonal matches with that of $\Delta$. Thus, 
\[
\Delta = \Delta^{+} + \Delta^{-} = \Delta^{+} + 
\Delta_{\mcl{S}}^{-} + \Delta_{\bar{\mcl{S}}}^{-}.
\]

We now make the following key observation. Since $\|\mbm{L}^{-}\|_1  = \|\mbm{L}_{\mcl{S}}^{-}\|_1$,
\begin{eqnarray}\label{eq:l1_norm_difference}
\|\mbm{L}+\Delta\|_{1,off} - \|\mbm{L}\|_{1,off} &=& \|\mbm{L}^{-}+\Delta^{-}\|_1 - \|\mbm{L}^{-}\|_1 
\nonumber\\
&=& \|\mbm{L}_{\mcl{S}}^{-} + \Delta_{\mcl{S}}^{-}\|_1 + \|\Delta_{\bar{\mcl{S}}}^{-}\|_1 - \|\mbm{L}^{-}\|_1  \nonumber\\
&\geq& \|\Delta_{\bar{\mcl{S}}}^{-}\|_1 - \|\Delta_{\mcl{S}}^{-}\|_1,
\end{eqnarray}
by triangle inequality. 

We also have 
\[
\|\Delta\|_1 = \|\Delta^{+}\|_1 + \|\Delta^{-}\|_1.
\]
Furthermore, 
\begin{equation}\label{eq:Delta_S_l1_F_norm_bound}
\|\Delta_{\mcl{S}}^{-}\|_1 \leq \sqrt{s} \|\Delta_{\mcl{S}}^{-}\|_F 
\leq \sqrt{s} \|\Delta \|_F.
\end{equation} 
and 
\begin{equation}\label{eq:Delta_plus_l1_F_norm_bound}
\|\Delta^{+}\|_1 \leq \sqrt{p}  \|\Delta^{+}\|_F \leq \sqrt{p}  \|\Delta \|_F,
\end{equation} 
so that 
\begin{equation}\label{eq:Delta_l1_F_norm_bound}
\|\Delta_{\mcl{S}}^{-}\|_1 + \|\Delta^{+}\|_1\leq (\sqrt{s}+\sqrt{p}) \|\Delta \|_F \leq 2\sqrt{s+p}
\|\Delta \|_F.
\end{equation}

Now, we focus on individual terms in the difference $\ell_0(\mbm{L}^*+\Delta) - \ell_0(\mbm{L}^*)$.  
First, 
\begin{eqnarray}\label{eq:trace_difference}
&&\frac{1}{2}\tr(\wh{\Sigma}(\theta_0 I_p + \mbm{L}^* +\Delta)^2) - \frac{1}{2}
\tr(\wh{\Sigma}(\theta_0 I_p + \mbm{L}^*))^2) \nonumber\\
&=& \tr((\theta_0 I_p + \mbm{L}^*)\wh{\Sigma}\Delta) + \frac{1}{2} \tr(\wh{\Sigma}\Delta^2).  
\end{eqnarray}
Next, by a second order Taylor series expansion (see Sec. 2 of \cite{RothmanBLZ2008}), for any symmetric $\Delta$, 
\begin{eqnarray}\label{eq:logdet_difference}
&& \log\det(\theta_0 I_p + \mbm{L}^*+\Delta) - \log\det(\theta_0 I_p + \mbm{L}^*) \nonumber\\
&=& \tr((\theta_0 I_p + \mbm{L}^*)^{-1}\Delta)  \nonumber\\
&&
 - \mbox{vec}(\Delta)^T \left[ \int_0^1 (1-t) (\theta_0 I_p + \mbm{L}^* + t\Delta)^{-1}\otimes  (\theta_0 I_p + \mbm{L}^* + t\Delta)^{-1} dt \right] \mbox{vec}(\Delta) \nonumber\\
&=& \tr((\theta_0 I_p + \mbm{L}^*)^{-1}\Delta)  - R(\Delta),
\end{eqnarray}
where $\otimes$ denotes Kronecker product between matrices. 

Let $\wt{\Sigma}^{1/2} := (\theta_0 I_p + \mbm{L}^*)\wh{\Sigma}$ and $\Sigma_*^{1/2} = 
(\theta_0 I_p + \mbm{L}^*)\Sigma_* = (\theta_0 I_p + \mbm{L}^*)^{-1}$. Then, by an application of Lemma \ref{lem:concen_quad_functional}, we can show that, given $c > 0$, there exists $C_1,C_2 > 0$ (depending on $\theta_0$ and $\|\mbm{L}^*\|$), such that 
\begin{eqnarray}
\mbm{P}\left( \max_{1\leq i\neq j\leq p} \left|(\wt{\Sigma}^{1/2})_{ij} - (\Sigma_*^{1/2})_{ij}\right| > C_1 \sqrt{\frac{\log p}{n}}\right) &\leq& p^2n^{-c} \label{eq:offdiag_covar_div_bound}\\
\mbm{P}\left(\max_{1\leq i\leq p} \left|(\wt{\Sigma}^{1/2})_{ii} - (\Sigma_*^{1/2})_{ii}\right| >  C_2 \sqrt{\frac{\log p}{n}}\right) &\leq& pn^{-c}.
\label{eq:diag_covar_div_bound}
\end{eqnarray}

Implication of (\ref{eq:offdiag_covar_div_bound}) and (\ref{eq:diag_covar_div_bound}) 
is that, we have, \textit{w.p. $\to 1$}, 
\begin{eqnarray}\label{eq:trace_diff_logdet_deriv_bound}
&& |\tr((\theta_0 I_p + \mbm{L}^*)\wh{\Sigma}\Delta) - \tr((\theta_0 I_p + \mbm{L}^*)^{-1}\Delta)| \nonumber\\
&=& |\tr((\theta_0 I_p + \mbm{L}^*)(\wh{\Sigma}-\Sigma_*)\Delta)| \nonumber\\
&=& |\tr((\wt{\Sigma}^{1/2} - \Sigma_*^{1/2})\Delta)| \nonumber\\
&\leq& |\sum_{i=1}^p ((\wt{\Sigma}^{1/2})_{ii} - (\Sigma_*^{1/2})_{ii}) \Delta_{ii}| +  |\sum_{i\neq j} ((\wt{\Sigma}^{1/2})_{ij} - (\Sigma_*^{1/2})_{ij}) \Delta_{ij}| \nonumber\\
&\leq& C_2 \sqrt{\frac{\log p}{n}} \|\Delta^{+}\|_1 + C_1 \sqrt{\frac{\log p}{n}} \|\Delta^{-}\|_1.
\end{eqnarray}

Notice also that 
\begin{equation}\label{eq:hat_Sigma_quad_Delta_lower_bound}
\tr(\wh{\Sigma}\Delta^2) \geq 0
\end{equation}
since both $\Delta^2 = \Delta \Delta^T$ and $\wh{\Sigma}$ are positive semidefinite matrices.

Combining (\ref{eq:trace_difference}), (\ref{eq:logdet_difference}), 
(\ref{eq:trace_diff_logdet_deriv_bound}), (\ref{eq:hat_Sigma_quad_Delta_lower_bound}) and (\ref{eq:l1_norm_difference}), we have, for any $\lambda > 0$, and any $\Delta \in \mcl{D}_p(\mbm{L}^*)$, 
\begin{eqnarray}\label{eq:G_lambda_prelim_bound}
G_\lambda(\Delta) &\geq& R(\Delta) + \tr((\theta_0 I_p +\mbm{L}^*)\wh{\Sigma}\Delta) \nonumber\\
&-& \tr((\theta_0 I_p 
+ \mbm{L}^*)^{-1}\Delta)  
+ \lambda \left( \|\Delta_{\bar{\mcl{S}}}^{-}\|_1 - \|\Delta_{\mcl{S}}^{-}\|_1 \right) \nonumber\\
&\geq& R(\Delta) - C_1 \sqrt{\frac{\log p}{n}} \|\Delta^{-}\|_1 - C_2 \sqrt{\frac{\log p}{n}} \|\Delta^{+}\|_1 
\nonumber\\
&+& \lambda \left( \|\Delta_{\bar{\mcl{S}}}^{-}\|_1 - \|\Delta_{\mcl{S}}^{-}\|_1 \right),
\end{eqnarray}
\textit{w.p. $\to 1$}.

As in \cite{RothmanBLZ2008}, we observe that eigenvalues of the Kronecker product of 
two positive semidefinite matrices are products of the eigenvalues of constituent matrices. 
Notice also that since $\|\Delta\|_F = r_n = o(1)$, for sufficiently large $n$, for all $t \in (0,1)$ the matrix 
$\theta_0 I_p + \mbm{L}^* + t\Delta$ is positive definite. Denoting by $\varphi_{\min}$ and $\varphi_{\max}$, the smallest and largest eigenvalues of a symmetric matrix, it therefore follows that
\begin{eqnarray}\label{eq:R_Delta_eigenbound}
&& \min_{0 < t < 1}\varphi_{\min}\left((\theta_0 I_p + \mbm{L}^* + t\Delta)^{-1}\otimes  (\theta_0 I_p + \mbm{L}^* + t\Delta)^{-1}\right) \nonumber\\
&=& \min_{0 < t < 1} \left(\varphi_{\max}(\theta_0 I_p + \mbm{L}^* + t\Delta)\right)^{-2} 
\nonumber\\
&\geq& (\theta_0 + \|\mbm{L}^*\| + \|\Delta\|_F)^{-2} = (\theta_0 + \|\mbm{L}^*\| + r_n)^{-2}
~\geq~ K_0, 
\end{eqnarray}
for $0 < K_0 \leq \frac{2}{3}(K_1 + K_2)^{-2}$, and for large enough $n$ (by \tb{A1} and \tb{A2}).
Consequently, for large enough $n$, for all $\Delta \in \mcl{D}_p(\mbm{L}^*)$,  
\begin{equation}\label{eq:logdet_Hessian_bound}
R(\Delta) \geq \frac{K_0}{2} \|\Delta\|_F^2.
\end{equation}

Now take 
\begin{equation}
\lambda = \lambda_n = C_3 \sqrt{\frac{\log p}{n}}
\end{equation}
for some $C_3 > C_1$. Then, defining $C_4 = 2(C_1 + C_2 + C_3)$, 
from (\ref{eq:G_lambda_prelim_bound}), it follows that \textit{w.p. $\to 1$}, 
uniformly in $\Delta \in \mcl{D}_p(\mbm{L}_0) \cap \mcl{B}_p^b(r_n)$,
\begin{eqnarray}
G_{\lambda_n}(\Delta) &\geq& \frac{K_0}{2} \|\Delta\|_F^2 + (C_3 - C_1) \sqrt{\frac{\log p}{n}} \|\Delta_{\bar{\mcl{S}}}^{-}\|_1 \
\nonumber\\
&-& (C_1+C_3)\sqrt{\frac{\log p}{n}} \|\Delta_{\mcl{S}}^{-}\|_1 
- C_2 \sqrt{\frac{\log p}{n}}\|\Delta^{+}\|_1 \nonumber\\
&\geq& \frac{K_0}{2} \|\Delta\|_F^2 - C_4 \sqrt{\frac{(s+p)\log p}{n}}  \|\Delta\|_F \nonumber\\
&=&  \|\Delta\|_F^2 \left(\frac{K_0}{2} - C_4 \sqrt{\frac{(s+p)\log p}{n}} \|\Delta\|_F^{-1}\right) \nonumber\\
&=& \|\Delta\|_F^2 \left(\frac{K_0}{2} - \frac{C_4}{M}\right) 
\end{eqnarray}
where the second inequality is by (\ref{eq:Delta_l1_F_norm_bound}) and 
noticing that $C_3 - C_1> 0$, and the last line follows since 
$\|\Delta\|_F = r_n = M \sqrt{\frac{(s+p)\log p}{n}}$. Now, if $M > 2C_4/K_0$, then 
the right hand side is strictly positive, uniformly in $\|\Delta\|_F^2$. 

This completes the proof of Theorem \ref{thm:consistency_hat_L_theta_0_known}.

\subsection{Proofs of Theorem \ref{thm:consistency_hat_L_theta_plugin} and  Lemma \ref{lem:eigen_rate}}

\noindent\tb{Proof of Theorem \ref{thm:consistency_hat_L_theta_plugin}:}
As before, we write $\Delta = \mbm{L}-\mbm{L}^*$. Further, let $u = \theta - \theta_0$ and 
let $\wh{u} = \wh{\theta} - \theta_0$. With $g_\lambda(\theta,\mbm{L})$ as defined
in (\ref{eq: GAR_loglike}), we have
\begin{eqnarray}\label{eq:loss_diff_plugin}
\tilde{G}_\lambda(\wh{u},\Delta) &:=& g_\lambda(\wh{\theta},\mbm{L}) - g_\lambda(\theta_0,\mbm{L}_0) \nonumber\\
&=&
\left(g_\lambda(\wh{\theta},\mbm{L}_0+\Delta) - g_\lambda(\theta_0,\mbm{L}_0+\Delta)  \right)	
\nonumber\\
&+& \left(g_\lambda(\theta_0,\mbm{L}_0+\Delta) - g_\lambda(\theta_0,\mbm{L}_0)  \right)	\nonumber\\
&=& T_1(\wh{u},\Delta) + T_2(\Delta).
\end{eqnarray}
Here,
\begin{eqnarray}\label{eq:T_2_Delta}
T_2(\Delta) &=& \tr((\theta_0 I_p + \mbm{L}_0)(\wh{\Sigma}-\Sigma_0)\Delta) + \frac{1}{2}\tr(\wh{\Sigma}\Delta^2)  \nonumber\\
&& + \lambda (\|\mbm{L}_0+\Delta\|_{1,off}- \|\mbm{L}_0\|_{1,off} ) + \wh{R}(\Delta), 
\end{eqnarray}
and 
\begin{eqnarray}
\label{eq:T_1_Delta}
T_1(\wh{u},\Delta) &=& 2\wh{u}~\tr(\Sigma_*\Delta) + \wh{u}~\tr((\wh{\Sigma} - \Sigma_*)\Delta) \nonumber\\
&-& \wh{u}^2\tr((\wh{\theta}I_p + \mbm{L}_0)^{-1}\Sigma_0\Delta)  
\end{eqnarray}
where $\wh{R}(\Delta)$ has the same form as $R(\Delta)$, with $\theta_0$ replaced by $\wh{\theta}$, and $\wh{R}(\Delta)$ is as in (\ref{eq:trace_difference}).

Comparing (\ref{eq:loss_diff_plugin}) with the analysis following (\ref{eq:trace_difference}) and (\ref{eq:logdet_difference}), we notice that the major difference 
is due to the term $T_1(\wh{u},\Delta)$. 

Specifically, as the analysis below shows, the terms that are linear in $\wh{u} = \wh{\theta}-\theta_0$ contribute to the change of rates.

As in the proof of Theorem \ref{thm:consistency_hat_L_theta_0_known}, we restrict attention to $\Delta$ belonging to the set $\mcl{B}_p^b(r_n)=\{\Delta \in \mathbb{R}^{p\times p}:\|\Delta\|_F=e_n\}$, where $r_n = M \sqrt{s+p}\lambda_n$, with suitably large positive $M$, and $\lambda_n$ as in the statement of Theorem \ref{thm:consistency_hat_L_theta_plugin}.

Similar to the analysis of $R(\Delta)$ in (\ref{eq:R_Delta_eigenbound}) and (\ref{eq:logdet_Hessian_bound}), and assumption 
\tb{B1}, we observe that for large enough $n$
(so that $|\wh{u}| = |\wh{\theta} -\theta_0| \leq \gamma_n$), \textit{w.p. $\to 1$}, 
\begin{equation}
\wh{R}(\Delta) \geq \frac{\wh{K}}{2} \|\Delta\|_F^2 
\end{equation}
where $0 < \wh{K} \leq (\theta_0 + \|\mbm{L}^*\|+r_n + \gamma_n)^{-2}$. Therefore, using bounds analogous to (\ref{eq:diag_covar_div_bound})--(\ref{eq:offdiag_covar_div_bound}), we observe that for large enough $n$, \textit{w.p. $\to 1$}, 
\begin{equation}\label{eq:T_2_Delta_LBD}
T_2(\Delta) \geq \frac{\wh{K}}{2} \|\Delta\|_F^2 
+ \lambda_n \left( \|\Delta_{\bar{\mcl{S}}}^{-}\|_1 - \|\Delta_{\mcl{S}}^{-}\|_1 \right)
- C_1 \sqrt{\frac{\log p}{n}} \|\Delta^{-}\|_1 - C_2 \sqrt{\frac{\log p}{n}} \|\Delta^{+}\|_1 
\end{equation}
for appropriate constants $C_1$ and $C_2$.

To control the contributions of the terms in $T_1(\wh{u},\Delta)$, we observe that 
\begin{eqnarray*}
|\tr(\Sigma_*\Delta)| &\leq& \|\Sigma_0\|_{\infty,\infty} \|\Delta\|_1 \\
|\tr((\wh{\theta}I_p + \mbm{L}_0)^{-1}\Sigma_0\Delta)| 
&\leq& \|(\wh{\theta}I_p + \mbm{L}_0)^{-1}\| ~\|\Sigma_0\|~\|\Delta\|_F ~=~
\frac{1}{\wh{\theta}\theta_0^2}~\|\Delta\|_F
\end{eqnarray*}
where $\|\Sigma\|_{\infty,\infty} = \max_{i,j}|\Sigma_{ij}|$. The second equality in the 
last line follows since $\mbm{L}^*$ is positive semidefinite. Note that $\|\Sigma_0\|_{\infty,\infty} \leq C$ for some $C>0$ by assumptions \tb{A1}--\tb{A2}.
Therefore, for suitable constants $C_3, C_4 > 0$, \textit{w.p. $\to 1$}, 
\begin{eqnarray}\label{eq:T_1_u_Delta_LBD}
T_1(\wh{u},\Delta) &\geq& - C_3 |\wh{u}| \|\Delta\|_1
- C_1 |\wh{u}| \sqrt{\frac{\log p}{n}} \|\Delta^{-}\|_1 
- C_2 |\wh{u}| \sqrt{\frac{\log p}{n}} \|\Delta^{+}\|_1
\nonumber\\
&&
~~~~ - C_4 |\wh{u}|^2 \|\Delta\|_F \nonumber\\
&\geq& - |\wh{u}| (C_3 + C_1  \sqrt{\frac{\log p}{n}}) \|\Delta^{-}\|_1 -|\wh{u}| (C_3 + C_2  \sqrt{\frac{\log p}{n}}) \|\Delta^{+}\|_1 \nonumber\\
&&~~~~
- C_4 |\wh{u}|^2 \|\Delta\|_F.
\end{eqnarray}

Combining (\ref{eq:T_2_Delta_LBD}) and (\ref{eq:T_1_u_Delta_LBD}), and using (\ref{eq:loss_diff_plugin}), we have that for 
suitable
constants $C_0,C_1,C_2,\wt{C}_3, C_4 >0$, \textit{w.p. $\to 1$}, 
\begin{eqnarray}
\tilde{G}_\lambda(\wh{u},\Delta) &\geq& C_0 \|\Delta\|_F^2 +  \lambda_n (\|\Delta_{\bar{\mcl{S}}}^{-}\|_1 - \|\Delta_{{\mcl{S}}}^{-}\|_1)
 - (C_1 \sqrt{\frac{\log p}{n}} +  \wt{C}_3|\wh{u}|) \|\Delta^{-}\|_1  \nonumber\\
 &-& \wt{C}_3 |\wh{u}| \|\Delta^{+}\|_1 - C_4 |\wh{u}|^2 \|\Delta\|_F.
\end{eqnarray}
Then, using the fact that $|\wh{u}| \leq \gamma_n$ for large enough $n$,  and $\|\Delta_{S}^{-}\|_1 \leq \sqrt{s}\|\Delta\|_F$ and $\|\Delta^{+}\|_1 \leq \sqrt{p}\|\Delta\|_F$, we have that for sufficiently large $C_4, C_5$ and $M_1$,  
\begin{eqnarray}\label{eq:riskdiff_plugin_lower}
	\tilde{G}_\lambda(\wh{u},\Delta) &\geq& C_0 \|\Delta\|_F^2  - C_5 (\sqrt{s+p}\lambda_n + \gamma_n^2)\|\Delta\|_F \nonumber\\
	&=& C_0 r_n^2\left(1- C_5 \frac{(\sqrt{s+p}\lambda_n + \gamma_n^2)}{r_n}\right),
\end{eqnarray}	
for all $\Delta \in \mcl{D}_p(\mbm{L}^*)\cap \mcl{B}_p^b(r_n)$.

Since $M$ can be chosen large enough that $C_5/M < 1/3$ and $\gamma_n^2\leq \lambda_n^2 
=O(r_n^2)$, it follows that for large enough $n$, the right hand side of (\ref{eq:riskdiff_plugin_lower}) is at least $(C_0/2)r_n^2$. This proves that there is a local minimum of the function $g_\lambda(\wh{\theta},\mbm{L})$ within a Frobenius norm radius of $r_n$ of $\mbm{L}^*$. By strong convexity of the function $g_\lambda(\wh{\theta},\mbm{L})$ with respect $\mbm{L}$, any local minimizer is a global 
minimizer, and hence the bound on $\wh{\mbm{L}}$ follows.

\noindent\tb{Proof of Lemma \ref{lem:eigen_rate}:} By \textit{Weyl's inequality}, and a bound on extreme eigenvalues of a Wishart-type matrix due to \cite{Vershynin2010}, \textit{w.p. $\to 1$}  
\begin{equation}\label{eq:eigen_max_bound}
|\lambda_{max}(\wh{\Sigma}) - \lambda_{max}(\Sigma^*)| \leq \|\wh{\Sigma} - \Sigma^*| \leq C (\sqrt{p/n}+p/n).
\end{equation}
On the other hand, with $\mathbf{P}$ and $\wh{\mathbf{P}}$ denoting the eigenprojection corresponding to the largest eigenvalue of $\Sigma^*$ and $\wh{\Sigma}$, respectively, we have, 
\[
(\Sigma^* -\lambda_{max}(\Sigma^*) \mathbb{I}_p) \wh{\mathbf{P}} 
= (\lambda_{max}(\wh{\Sigma}) - \lambda_{max}(\Sigma^*))\wh{\mathbf{P}} - (\wh{\Sigma} - \Sigma^*)\wh{\mathbf{P}}
\]
Pre-multiplying by $\mathbf{P}$ and changing, 
\begin{eqnarray*}
(\lambda_{max}(\wh{\Sigma}) - \lambda_{max}(\Sigma^*))\mathbf{P}
&=& \mathbf{P} (\wh{\Sigma} - \Sigma^*)\mathbf{P}
+ \mathbf{P} (\wh{\Sigma} - \Sigma^*)(\wh{\mathbf{P}}-\mathbf{P}) \nonumber\\
&&
+ (\lambda_{max}(\wh{\Sigma}) - \lambda_{max}(\Sigma^*))\mathbf{P}(\wh{\mathbf{P}} - \mathbf{P}).
\end{eqnarray*}
Now, taking trace on both sides, using (\ref{eq:eigen_max_bound}), and a version of the 
\textit{Davis-Kahan inequality} on eigenprojections
\cite{YuWS2015}, together with the non-vanishing eigengap between the largest and second largest eigenvalue of $\Sigma^*$,
we obtain the result.

\subsection{Auxiliary lemmas}

\subsubsection*{Concentration of trace of a sample covariance matrix}
\label{subsec:concentration_trace}

\begin{lemma}\label{lem:trace_covar_diff}
Let $\mbf{B}$ be any $p \times p$ symmetric matrix. Also, let $Z_{ij}$'s be i.i.d. $\sigma$-sub-Gaussian with 
mean 0 and variance 1.Let $\mbf{B}$ be any $p \times p$ symmetric matrix. Also, let $Z_{ij}$'s be i.i.d. $\sigma$-sub-Gaussian with 
mean 0 and variance 1.  Let $c>0$ be arbitrary and $n$  and $p$ satisfy  $n > (c/b_0) \log p$, for some universal constant $b_0 >0$. Then there exists a $c_0>0$ such that
\begin{equation}\label{eq:trace_covar_diff_tight}
\mbm{P}\left(|\tr(\mbf{B}(\wh{\Sigma}-\Sigma_0))
> c_0 \|\Sigma_0\| \|\mbf{B}\|_F \sqrt{\frac{\log p}{n}}\right) \leq 2p^{-c}.
\end{equation}
\end{lemma}

\vskip.1in
\noindent\textbf{Proof of Lemma \ref{lem:trace_covar_diff}:} First let
$\mcl{Z} =$ vec$(\mbf{Z}^T)$, a $pn\times 1$ block vector, with $i$-th block equal to the $i$-th row of $\mbf{Z}$, for $i=1,\ldots,p$. Then, 
we can express $\tr(\mbf{B}\wh{\Sigma})$ as $\mcl{Z}^T\mbf{A} \mcl{Z}$, where $\mbf{A}$ is the $np \times np$ matrix given by 
$(\Sigma_0^{1/2}\mbf{B}\Sigma_0^{1/2}) \otimes I_n$, where $\otimes$ denotes the Kronecker product. 

Then
\[
\tr(\mbf{B}(\wh{\Sigma}-\Sigma_0)) = \frac{1}{n}\left(\mcl{Z}^T\mbf{A} \mcl{Z} - 
\mbm{E}(\mcl{Z}^T\mbf{A} \mcl{Z})\right). 
\]
So, by \textit{Hanson-Wright inequality}, we have that for any $t> 0$, 
\begin{eqnarray}
	&&\mbm{P}(|\tr(\mbf{B}(\wh{\Sigma}-\Sigma_0))|>t)=
	\mbm{P}\left(|\mcl{Z}^T\mbf{A} \mcl{Z} - 
	\mbm{E}(\mcl{Z}^T\mbf{A} \mcl{Z})|>nt\right) \nonumber\\
	&\leq& 2\exp\left\{
	-b_0  \min\left(\frac{n^2t^2}{\sigma^4\|\mbf{A}\|_F^2},
	~\frac{nt}{\sigma^2\|\mbf{A}\|}\right) 
	\right\},
\end{eqnarray}
for a constant $b_0>0$. Notice that  $\|\mbf{A}\| = \|\Sigma_0^{1/2} \mbf{B}\Sigma_0^{1/2}\| \leq \|\Sigma_0\| \|\mbf{B}\|$, and 
\[
\|\mbf{A}\|_F^2 = n \|\Sigma_0^{1/2} \mbf{B}\Sigma_0^{1/2}\|_F^2 
\leq n \|\Sigma_0\|^2 \|\mbf{B}\|_F^2.
\]
Now, set $c_0 =\sigma_0^2 \sqrt{c/b_0}$. 
Since $n > (c/b_0) \log p$, we can use the above bound for
$\|\mbf{A}\|_F^2$ and set  $t = c_0  \|\Sigma_0\|\|\mbf{B}\|_F \sqrt{\log p/n}$ to  obtain (\ref{eq:trace_covar_diff_tight}).

\subsubsection*{Concentration of quadratic functionals}\label{subsec:concentration_quadratic}

\begin{definition}
We say that an $n\times p$ random matrix $\mbf{X}$ is a zero mean, sub-Gaussian matrix with parameters $(\Gamma,\sigma^2)$ if 
\begin{itemize}
\item[(a)] each row $X_i^T \in \mbm{R}^p$ is sampled independently from a zero-mean distribution with covariance matrix $\Gamma$, and 
\item[(b)] for any unit vector $\mbf{u} \in \mbm{R}^p$, the random variable
$\mbf{u}^T X_i$ is a sub-Gaussian random variable with parameter $\sigma$.
\end{itemize}
\end{definition}

The following lemma on fluctuations of quadratic forms involving sample covariance matrices for sub-Gaussian random variables is from \cite{lohw2012}.

\begin{lemma}\label{lemma:sub_Gaussian_quadratic}
If $\mbf{X} \in \mbm{R}^{n \times p}$ is a zero-mean sub-Gaussian matrix with parameters $(\Gamma,\sigma^2)$, we have, for any $t > 0$, and any unit vector $\mbf{v} \in \mbm{R}^p$, 
\begin{equation}\label{eq:sub_Gaussian_quadratic}
\mbm{P}\left(|\|\mbf{X}\mbf{v}\|^2 - \mbm{E}(\|\mbf{X}\mbf{v}\|^2)| \geq nt\right) \leq 2 \exp\left(-c n \min\{\frac{t^2}{\sigma^4},\frac{t}{\sigma^2}\}\right)
\end{equation}	
for some universal constant $c > 0$.
\end{lemma}

\begin{lemma}\label{lem:concen_quad_functional}
Suppose that $\mbf{Z}$ is $p \times n$ matrix consisting of i.i.d. $N(0,1)$ entries. Then, there exist constants $c_1,c_2, \delta > 0$ such that for for any pair of unit vectors $\mbf{u}, \mbf{v} \in \mbm{R}^p$, we have 
\begin{equation}\label{eq:concen_quad_functional}
\mbm{P}\left(  \left| \mbf{u}^T \left(\frac{1}{n} \mbf{Z}\mbf{Z}^T \right) \mbf{v}  - \mbf{u}^T \mbf{v} \right| > t \right) \leq c_1 \exp(-c_2 n t^2) \qquad \mbox{for} \qquad 0 < t < \delta. 
\end{equation}	
\end{lemma}

\subsubsection*{Proof of Lemma \ref{lem:concen_quad_functional}:}
Let $P_{\mbf{u}} = \mbf{u}\mbf{u}^T$ and let $U$ be a $p\times (p-1)$ matrix such that $UU^T = I_p - P_{\mbf{u}}$. Then, writing $\mbf{v} = (\mbf{u}^T \mbf{v}) \mbf{u} + (I_p - P_{\mbf{u}})\mbf{v}$, we can decompose 
\[
\mbf{u}^T \left(\frac{1}{n} \mbf{Z}\mbf{Z}^T \right) \mbf{v}  - \mbf{u}^T \mbf{v} = (\mbf{u}^T \mbf{v}) \left(\frac{1}{n} \|\mbf{Z}^T\mbf{u}\|^2 - 1\right) + \frac{1}{n} \langle \mbf{Z}^T\mbf{u}, (\mbf{Z}^T U) \wt{\mbf{v}}\rangle,
\]
where $\wt{\mbf{v}} = U^T \mbf{v}$. Key to completing the proof is to notice that $\mbf{Z}^T \mbf{u}$ and $\mbf{Z}^T U$ are independent and with i.i.d. $N(0,1)$ entries.

\section{Additional simulation results}
\label{sec:appendix_simulation}

\subsection{Denser graphs}

 In this setting, we raised the edge probability to $\frac{5}{p}$, where $p$ is the signal dimension.  This leads to graphs with $261, 694, 1238$ edges for $p=100,250, 500$, respectively. Moreover, there is no self-loop. Given the graph, data generation and results evaluation follow those in Section \ref{subsec:simu_data_generation} and Section \ref{subsec:simu_eval}, respectively.

\begin{table}[H]
    \centering
    \begin{tabular}{|c|c|c|c|c|c|c|c|}
    \hline
     \multicolumn{2}{|c|}{\bf Setting} & $\theta_0$  & $\mathbf{v}_0$  &  $\mbb{L}$   & {\bf Power} & {\bf FDR} & {\bf F1} \\
		\multicolumn{2}{|c|}{}& Error &  Error  &Error & & &\\
		\hline
 \multirow{3}{*}{$p=100$} & $n=100$ & .570 & .140 & .231 & .420 & .108 & .571\\
    & $n=250$ & .013 & .043 & .031 & .807 & .076 & .862\\ 
     & $n=500$ & .003 & .020 & .011 & .925 & .021 & .951\\
     \cline{1-8}
    \multirow{3}{*}{$p=250$} & $n=100$ & .792 & .166 & .288 & .353 & .223 & .485\\
    & $n=250$ & .199 & .096 & .093 & .613 & .033 & .750\\ 
      & $n=500$ & .005 & .030 & .015 & .894 & .034 & .929\\ 
       \cline{1-8}
     \multirow{3}{*}{$p=500$} & $n=100$ & .307 & .146 & .173 & .295 & .438 & .387\\
     & $n=250$ & .220 & .110 & .091 & .651 & .028 & .780\\ 
      & $n=500$ & .046 & .039 & .020 & .875 & .010 & .929\\ 
    \hline
    \end{tabular}
    \caption{Edge Probability $\frac{5}{p}$: Results for the eBIC-selected GAR model.}
    \label{tab:eBIC_5p}
\end{table}

\begin{table}[H]
    \centering
    \begin{tabular}{|c|c|c|c|c|}
    \hline
 \multicolumn{2}{|c|}{\bf Setting} & $\wh{\theta}_0^{(ini)}$ & Step 1 $\wh{\mbb{L}}_{\lambda}$ & {\bf Oracle} $\wh{\mbb{L}}$ \\ 
  \multicolumn{2}{|c|}{}  & Error & Error & Error \\ 
    \hline
    \multirow{3}{*}{$p=100$} & $n=100$ & .008 & .126 & .026 \\ 
    & $n=250$ & .003 & .072 & .010 \\ 
    & $n=500$ & .002 & .039 & .005 \\ 
    \hline
    \multirow{3}{*}{$p=250$} & $n=100$ & .030 & .132 & .028 \\ 
    & $n=250$ & .007 & .087 & .011 \\ 
    & $n=500$ & .003 & .052 & .006 \\ 
    \hline
    \multirow{3}{*}{$p=500$} & $n=100$ & .087 & .137 & .025 \\ 
    & $n=250$ & .025 & .086 & .010 \\ 
    & $n=500$ & .008 & .051 & .005 \\ 
    \hline
    \end{tabular}
    \caption{Edge Probability $\frac{5}{p}$: Results for the Step 1 GAR estimator  ($\lambda=\sqrt{\log p/n}$) and  the oracle GAR estimator.}
    \label{tab:5_p_oracle_step1}
\end{table}

\subsection{Graphs with self-loops}

In this setting, the edge-probability is $2/p$ as in the ``baseline simulation",  whereas  the self-loop probability  is $\frac{40}{p}$.  For signal dimension $p=100, 250, 500$, this resulted in a graph with $35$, $45$, and $47$  self-loops, respectively.  

\begin{table}[H]
    \centering
    \begin{tabular}{|c|c|c|c|c|c|c|c|}
    \hline
     \multicolumn{2}{|c|}{\bf Setting} & $\theta_0$  & $\mathbf{v}_0$  &  $\mbb{L}$   & {\bf Power} & {\bf FDR} & {\bf F1} \\
		 \multicolumn{2}{|c|}{}& Error &  Error  &Error & & & \\
		\hline
		\
 \multirow{3}{*}{$p=100$} & $n=100$ & .008 & .078 & .024 & .938 & .070 & .934\\
    & $n=250$ & $<.001$ & .028 & .007 & .996 & .039 & .978 \\ 
    & $n=500$ & $<.001$ & .018 & .003 & $>.999$ & .014 & .993\\
    \cline{1-8}
    \multirow{3}{*}{$p=250$} & $n=100$ & .032 & .079 & .037 & .872 & .087 & .892 \\
    & $n=250$ & .005 & .028 & .008 & .980 & .012 & .976\\ 
      & $n=500$ & $<.001$ & .018 & .004 & $>.999$ & .020 & .990 \\ 
         \cline{1-8}
     \multirow{3}{*}{$p=500$} & $n=100$ & .020 & .089 & .031 & .901 & .147 & .876\\
     & $n=250$ & .013 & .037 & .008 & .982 & .040 & .971 \\ 
      & $n=500$ & .008 & .020 & .003 & .998 & .014 & .992\\ 
    \hline 
    \end{tabular}
    \caption{Edge probability $\frac{2}{p}$ and self-loop probability $\frac{40}{p}$: Results for the eBIC-selected GAR model.}
    \label{tab:eBIC_selfloop}
\end{table}

\begin{table}[H]
    \centering
    \begin{tabular}{|c|c|c|c|c|}
    \hline
    \multicolumn{2}{|c|}{\bf Setting} & $\wh{\theta}_0^{(ini)}$ & Step 1 $\wh{\mbb{L}}_{\lambda}$ & {\bf Oracle} $\wh{\mbb{L}}$ \\ 
    \multicolumn{2}{|c|}{}  & Error & Error & Error \\ 
    \hline
    \multirow{3}{*}{$p=100$} & $n=100$ & .086 & .067 & .012 \\
    & $n=250$ & .039 & .029 & .005 \\ 
    & $n=500$ & .020 & .015 & .002 \\ 
    \hline
    \multirow{3}{*}{$p=250$} & $n=100$ & .120 & .086 & .013 \\ 
    & $n=250$ & .052 & .038 & .005 \\ 
    & $n=500$ & .025 & .020 & .003 \\ 
    \hline
    \multirow{3}{*}{$p=500$} & $n=100$ & .238 & .085 & .013 \\ 
    & $n=250$ & .130 & .038 & .005 \\ 
    & $n=500$ & .075 & .020 & .003 \\ 
    \hline
    \end{tabular}
    \caption{Edge probability $\frac{2}{p}$ and self-loop probability $\frac{40}{p}$: Results under Step 1 fitted GAR model ($\lambda=\sqrt{\log p/n}$) and the oracle GAR estimator.}
    \label{tab:selfloop_oracle_step1}
\end{table}

\subsection{Goodness-of-fit measure and eBIC results}
Here, we report the minimum and maximum (across $100$ replicates) goodness-of-fit measure (\ref{eq:goodness-of-fit}) while fitting a GAR model to the data when data are generated according to a GAR model and generated according to a power-law GGM (non-GAR model), respectively. As we can see from Table \ref{tab:fit_parbootstrap}, this measure is effective when $p \leq n$: $GF \approx 1$ (indicating  the GAR model is a good fit to the data) when the true data generating model is indeed GAR, whereas $GF \approx 0$ (indicating the GAR model is not a good fit to the data) when the true data generating model is non-GAR. However, when $p>n$, this measure does not work anymore as it is always $0$ no matter  the data generating model is GAR or not. Therefore, we also examine the eBIC score for the selected model when fitting the GAR model and the \textit{glasso} model to the data, respectively. From Table \ref{tab:GAR_vs_glasso_ebic_results}, it can be seen that, when the true data generating model is GAR, the GAR eBIC selected model always has smaller eBIC score compared to the corresponding \textit{glasso} eBIC selected model. On the other hand, when the true data generating model is non-GAR, the \textit{glasso} models have smaller eBIC scores.  This implies that, in practice, particularly when $p>n$, we can use the eBIC score to decide which of the GAR and \textit{glasso} models provides  a better fit to the data. 

\begin{table}[H]
    \centering
    \begin{tabular}{|c|c|c|c|c|}
    \hline
    \multicolumn{3}{|c|}{\bf Setting} & \bf Min GF & \bf Max GF \\ 
    \hline
    \multirow{9}{*}{True model: GAR} & \multirow{3}{*}{$p=100$} 
    & $\mathbf{n=100}$ & .880 & 1 \\
     && $\mathbf{n=250}$ & 1 & 1 \\
         & & $\mathbf{n=500}$ & 1 & 1 \\
         \cline{2-5}
         & \multirow{3}{*}{$p=250$} & $n=100$ & 0 & 0 \\ 
         & & $\mathbf{n=250}$ & 1 & 1 \\
         & & $\mathbf{n=500}$ & 1 & 1 \\ 
         \cline{2-5}
         & \multirow{3}{*}{$p=500$} & $n=100$ & 0 & 0 \\ 
         & & $n=250$ & 0 & 0 \\ 
         & & $\mathbf{n=500}$ & 1 & 1 \\ 
         \hline
         \multirow{9}{*}{True model: Non-GAR} & \multirow{3}{*}{$p=100$} & $\mathbf{n=100}$ & 0 & 0 \\
       & & $\mathbf{n=250}$ & 0 & 0 \\
         & & $\mathbf{n=500}$ & 0 & .020 \\ 
         \cline{2-5}
         & \multirow{3}{*}{$p=250$} & $n=100$ & 0 & 0 \\ 
         & & $\mathbf{n=250}$ & 0 & 0\\ 
         & & $\mathbf{n=500}$ & 0 & 0 \\
         \cline{2-5}
         & \multirow{3}{*}{$p=500$} & $n=100$ & 0 & 0 \\
         & & $n=250$ & 0 & 0 \\ 
         & & $\mathbf{n=500}$ &  0 & 0 \\ 
         \hline
    \end{tabular}
    \caption{Parametric bootstrap goodness-of-fit measure when fitting the GAR model.}
    \label{tab:fit_parbootstrap}
\end{table}

\begin{table}[H]
    \centering
    \begin{tabular}{|c|c|c|c|c|}
    \hline
    \multicolumn{3}{|c|}{\bf Setting} & \textbf{Fitted Model} & \textbf{eBIC Score} \\ 
    \hline
    \multirow{18}{*}{True model: GAR}
    &\multirow{6}{*}{$p=100$} & \multirow{2}{*}{$n=100$} & GAR & 10322 \\ 
    && & glasso & 10472 \\
         \cline{3-5}
    && \multirow{2}{*}{$n=250$} & GAR & 23467 \\ 
    && & glasso & 24090 \\ 
        \cline{3-5}
    && \multirow{2}{*}{$n=500$} & GAR & 46147 \\ 
    && & glasso & 47078 \\ 
     \cline{2-5}
    &\multirow{6}{*}{$p=250$} & \multirow{2}{*}{$n=100$} & GAR & 25261 \\ 
    && & glasso & 25293 \\ 
         \cline{3-5}
    && \multirow{2}{*}{$n=250$} & GAR & 57286 \\ 
    && & glasso & 57825 \\ 
      \cline{3-5}
    && \multirow{2}{*}{$n=500$} & GAR & 108729 \\ 
    && & glasso & 110824 \\ 
        \cline{2-5}
    &\multirow{6}{*}{$p=500$} & \multirow{2}{*}{$n=100$} & GAR & 52982 \\ 
    &&& glasso & 53452 \\
          \cline{3-5}
    && \multirow{2}{*}{$n=250$} & GAR & 120422 \\ 
    && & glasso & 121344 \\ 
          \cline{3-5}
    && \multirow{2}{*}{$n=500$} & GAR& 232114 \\
    && & glasso& 234093\\\hline
    
    \multirow{18}{*}{True model: Non-GAR}
    &\multirow{6}{*}{$p=100$} & \multirow{2}{*}{$n=100$} & GAR & 28861 \\ 
    && & glasso & 28317 \\
         \cline{3-5}
    && \multirow{2}{*}{$n=250$} & GAR & 70422 \\ 
    && & glasso & 69523 \\ 
        \cline{3-5}
    && \multirow{2}{*}{$n=500$} & GAR & 140201 \\ 
    && & glasso & 138160 \\ 
     \cline{2-5}
    &\multirow{6}{*}{$p=250$} & \multirow{2}{*}{$n=100$} & GAR & 72645 \\ 
    && & glasso & 70486 \\ 
         \cline{3-5}
    && \multirow{2}{*}{$n=250$} & GAR & 177861 \\ 
    && & glasso & 173318 \\ 
      \cline{3-5}
    && \multirow{2}{*}{$n=500$} & GAR & 350379 \\ 
    && & glasso & 342947 \\ 
        \cline{2-5}
    &\multirow{6}{*}{$p=500$} & \multirow{2}{*}{$n=100$} & GAR & 146017 \\ 
    &&& glasso & 141066 \\
          \cline{3-5}
    && \multirow{2}{*}{$n=250$} & GAR & 356902 \\ 
    && & glasso & 346834 \\ 
          \cline{3-5}
    && \multirow{2}{*}{$n=500$} & GAR & 706586 \\
    && & glasso & 687097 \\
    \hline
    \end{tabular}
    \caption{ eBIC score of the selected GAR model and of the  selected \textit{glasso} model.}
    \label{tab:GAR_vs_glasso_ebic_results}
\end{table}

\section*{Funding}
The authors gratefully acknowledge the following support: NSF-DMS-1915894.

\bibliographystyle{agsm}
\bibliography{gar}

\end{document}